\documentclass[%
 reprint,
 amsmath,amssymb,
 aps,
]{revtex4}

\usepackage{graphicx}
\usepackage{dcolumn}
\usepackage{bm}
\usepackage{subfig}
\usepackage{hhline}

\begin{document}

\preprint{APS/123-QED}

\def\mean#1{\left< #1 \right>} 
\title{On solving the Thomas Bargman-Michel-Telegdi equation using the Bogoliubov Krylov method of averages and the calculation of the Berry phases}

\author{M. Haj Tahar and C. Carli}
\affiliation{%
CERN, Geneva, Switzerland
}%

\begin{abstract}
Several proposals aimed at measuring the Electric Dipole Moment (EDM) for charged particles require very precise simulations and understanding of the systematic errors that can contribute to a spin buildup mimicking the EDM signal to be detected. In what follows, one used the Bogoliubov-Krylov-Mitropolski method of averages to solve the T-BMT equation and calculate the Berry phases arising for a proton EDM storage ring. The formalism employed proved to be particularly useful to determine the evolution of the spin at the observation point, i.e. at the location of the polarimeter. Several selected cases of lattice imperfections were simulated and benchmarked with the analytical estimates. This allowed the proof of the convergence of the numerical simulations and helped gain better understanding of the systematic errors. 
\end{abstract}

\pacs{Valid PACS appear here}
\maketitle



\section{Introduction}
A storage ring represents a new attractive method to search and measure the electric dipole moment of the proton (or other charged particles) by using polarized particles at their magic momentum \cite{farley}. However, to reach the sensitivity level required for such a measurement, it is crucial to understand and mitigate the systematic errors that can yield a fake signal mimicking the EDM one. The objective of this paper is to contribute some better understanding regarding that matter: starting from the spin precession equation, we will establish the formalism and all necessary quantities to compute the spin evolution in a storage ring. Then, using a perturbation method, an approximate solution to this equation will be derived and benchmarked with numerical simulations in MATHEMATICA. Then, using the tracking code BMAD, we will investigate the impact of several imperfections on the spin buildup for a fully electrostatic ring lattice.

\section{Frozen Spin}
\begin{figure}
\centering 
\includegraphics*[width=10cm]{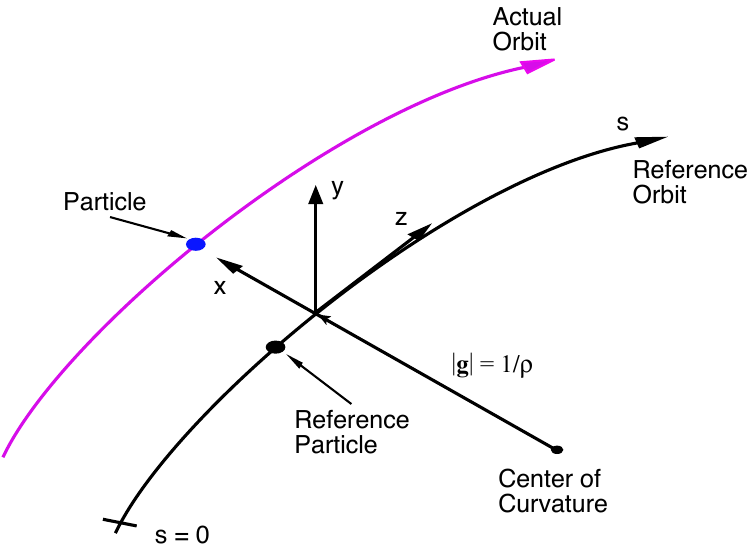}
\caption{The local reference coordinate system in BMAD \cite{bmad}. The curvature vector $\bm{g}$ lies in the median plane of the accelerator (sketch from the BMAD manual).}
\label{fig:local_coords}
\end{figure}

The Thomas Bargman-Michel-Telegdi (T-BMT) equation gives the precession rate of the angle between the spin and momentum vectors of a relativistic particle in the presence of electromagnetic fields \cite{thomas}\cite{bargman}:

\begin{eqnarray}
\dfrac{d\textbf{s}}{dt}= \left(\bm{\Omega_{BMT}}+\bm{\Omega_{EDM}}\right) \times \bm{s} \label{bmt}
\end{eqnarray}
where 
\begin{eqnarray}
\bm{\Omega_{BMT}} &=& -\dfrac{e}{mc} \left[ \left(a+\dfrac{1}{\gamma} \right) c \bm{B} - \dfrac{a\gamma c}{\gamma +1} (\bm{\beta . B})\bm{\beta} - \left(a+\dfrac{1}{\gamma + 1} \right) \bm{\beta} \times \textbf{E} \right] \\
 &=& -\dfrac{e}{mc} \left[ \left(a+\dfrac{1}{\gamma} \right) c \bm{B}_{\perp} + \dfrac{(1+a)c}{\gamma} \bm{B}_{\|} -\left(a+\dfrac{1}{\gamma + 1} \right) \bm{\beta} \times \textbf{E} \right] \nonumber
\end{eqnarray}
is the precession vector due to the particle's magnetic moment and
\begin{eqnarray}
\bm{\Omega_{EDM}} = - \dfrac{e}{mc} \dfrac{\eta}{2} \left[ \bm{E} - \dfrac{\gamma}{\gamma + 1} (\bm{\beta}.\bm{E})\bm{\beta} + c \bm{\beta} \times \bm{B} \right]
\end{eqnarray}
is the precession vector due the particle's finite electric dipole moment \cite{edm}. The charged particle considered throughout this paper is the proton. \\
The above equation describes the spin changes with the time $t$ in the laboratory frame. \textbf{s} is the classical spin vector expressed in the rest frame of the particle while $\bm{B}$, $\bm{E}$ are the magnetic and electric field vectors expressed in the laboratory frame. $\bm{B}_{\perp}$ and $\bm{B}_{\|}$ are the components perpendicular and parallel to the particle's momentum. $a$ is the particle's anomalous gyro-magnetic factor and $\gamma$ is the relativistic gamma factor. \\
In accelerator coordinates, i.e. in a frame co-moving with the reference particle (curvilinear coordinates system), the above equation can be re-written in the following way \cite{hoff}:
\begin{eqnarray}
\dfrac{d\textbf{s}}{ds}= \left( \dfrac{dt}{ds}\bm{\Omega_{BMT}} - \bm{\kappa} \times \bm{e_l} \right)\times \bm{s} \label{bmt2}
\end{eqnarray}
where $s$ is the arc length of the reference particle, $\bm{e_l}$ or $\bm{e_z}$ is the unit vector in the longitudinal direction and $\bm{\kappa}$ is the curvature vector as illustrated in fig \ref{fig:local_coords}. 

For simplicity, we will first assume that there are no magnetic fields and no EDM effects, i.e. $\eta=0$ and $\bm{B}=\bm{0}$.  \\
Defining $\rho$ as the radius of curvature of the reference orbit, one can write:
\begin{eqnarray}
\dfrac{dt}{ds} = \dfrac{1+x/\rho}{\beta_l c} \hspace{2mm};\hspace{2mm} e E_r^b = -\dfrac{\gamma_m E_0{\beta_m}^2}{\rho} \hspace{2mm};\hspace{2mm} \bm{\kappa} = \dfrac{1}{\rho} \bm{u_x} \label{rho}
\end{eqnarray}
where $\beta_m$ and $\gamma_m$ are the Lorentz factors at the magic energy, $E_0$ the energy of mass of the proton and $E_r^b$ is the radial electric field of the electrostatic bends for the ideal particle. Thus, in the following formalism, the spin buildup is measured with respect to the planar reference orbit \cite{silenko}. This will be the reference frame that we will use both for the analytical estimates as well as the numerical simulations. \\
In matrix notation, Eq. (\ref{bmt2}) writes: 
\begin{align}
\renewcommand\arraystretch{1.5}
    \begin{pmatrix}
        ds_{r}/ds \\
        ds_{y}/ds \\
        ds_{l}/ds
    \end{pmatrix} 
&=
\dfrac{e}{mc} \left[a+\dfrac{1}{\gamma+1} \right] \dfrac{dt}{ds}
\renewcommand\arraystretch{1.5}
    \begin{pmatrix}
        \left[(\beta_l E_r - \beta_r E_l) + \dfrac{mc}{e} \left(a + \dfrac{1}{\gamma +1} \right)^{-1} \dfrac{1}{\rho} ds/dt  \right] s_l + (\beta_y E_r -\beta_r E_y) s_y \\
        (\beta_r E_y - \beta_y E_r)s_r + (\beta_l E_y -\beta_y E_l) s_l \\
        (\beta_y E_l - \beta_l E_y)s_y -  \left[(\beta_l E_r - \beta_r E_l) + \dfrac{mc}{e} \left(a + \dfrac{1}{\gamma +1} \right)^{-1} \dfrac{1}{\rho} ds/dt  \right] s_r
    \end{pmatrix}   \label{dsds}
\end{align}
where the subscripts $r$ or $x$ denote equivalently the horizontal transverse coordinate pointing towards outer radii, $y$ denotes the vertical transverse coordinate orthogonal to the median plane of the unperturbed particle motion and $l$ or $z$ the longitudinal coordinate in the Frenet Serret frame. 
Differentiating with respect to time, one obtains:

\begin{align}
\renewcommand\arraystretch{1.8}
    \begin{pmatrix}
        ds_{r}/dt \\
        ds_{y}/dt \\
        ds_{l}/dt
    \end{pmatrix} 
&=
\dfrac{e}{mc}
\renewcommand\arraystretch{2.0}
    \begin{pmatrix}
        \left( \left[a+\dfrac{1}{\gamma+1} \right] ( \beta_l E_r - \beta_r E_l) + \dfrac{mc}{e} \dfrac{1}{\rho} \dfrac{ds}{dt} \right) s_l + \left[a+\dfrac{1}{\gamma+1} \right](\beta_y E_r -\beta_r E_y) s_y \\
        \left[a+\dfrac{1}{\gamma+1} \right] (\beta_r E_y - \beta_y E_r)s_r + \left[a+\dfrac{1}{\gamma+1} \right]  (\beta_l E_y -\beta_y E_l) s_l \\
        \left[a+\dfrac{1}{\gamma+1} \right] (\beta_y E_l - \beta_l E_y)s_y -  \left( \left[a+\dfrac{1}{\gamma+1} \right] ( \beta_l E_r - \beta_r E_l) + \dfrac{mc}{e} \dfrac{1}{\rho} \dfrac{ds}{dt} \right) s_r
    \end{pmatrix} 
    \\
&=      
\renewcommand\arraystretch{1.5}
    \begin{pmatrix}
        0 & -\Omega_l & \Omega_y \\
        \Omega_l & 0 & -\Omega_r \\
        -\Omega_y & \Omega_r & 0
    \end{pmatrix}    
    \begin{pmatrix}
        s_{r} \\
        s_{y} \\
        s_{l}
    \end{pmatrix}    \label{matrix} 
\end{align}
which writes in the general form $d\bm{S}/dt=\bm{M} \bm{S}$. Now, we compute the spin precession components ($\Omega_r$, $\Omega_y$, $\Omega_l$) in order to express them as a function of the accelerator coordinates $(x, x', y, y', z, \Delta p/p)$ in the Frenet Serret system (where $'=d/ds$).
For that, one has to express $(\beta_r, \beta_y, \beta_l)$ as a function of $(x, x', y, y', z, \Delta p/p)$.

\subsection{Calculation of the Lorentz factors}
Given that $\bm{p}= \gamma m c \bm{\beta}$ and $E = \gamma m c^2 = \gamma E_0$ where $E_0$ is the mass energy, then the normalized particle velocity is obtained:
\begin{eqnarray}
\bm{\beta} = \dfrac{\bm{p} c}{E}  \label{beta_form}
\end{eqnarray}
Recalling that $E^2 = p^2c^2 + E_0^2$, it results that for a particle with a momentum offset $\Delta p$ and slope \mbox{$\Delta p/p=(p-p_{ref})/p_{ref}$:}
\begin{eqnarray}
\beta = \dfrac{1+\dfrac{\Delta p}{p}}{\left[\left(1 + \dfrac{\Delta p}{p} \right)^2 + \left(\dfrac{E_0}{p_{ref} c} \right)^2  \right]^{1/2}}
\end{eqnarray}
In electrostatic elements, the change in energy implies a change in the particle momentum, i.e. a variation of $\Delta p/p$.
It results from Eq. (\ref{beta_form}) that:
\begin{eqnarray}
\gamma = \dfrac{p_{ref} c \left(1 + \dfrac{\Delta p}{p} \right)}{\beta E_0} = \left[1 + \left(\dfrac{p_{ref} c}{E_0}\right)^2 \left(1 +\dfrac{\Delta p}{p} \right)^2 \right]^{1/2} \label{gamma}
\end{eqnarray}
In accordance with BMAD \cite{bmad}, which has been used for the simulation presented in this paper, the phase space momenta are defined as:
\begin{eqnarray}
p_{x,y} = \dfrac{P_{x,y}}{P_{ref}} = \dfrac{\beta_{r,y} \gamma}{\beta_{ref} \gamma_{ref}} = \dfrac{\beta_{r,y}}{\beta} \left(1 + \dfrac{\Delta p}{p} \right)
\end{eqnarray}
and 
\begin{eqnarray}
\beta_r &=& \dfrac{1}{c}\dfrac{dx}{dt} = \dfrac{1}{c}\dfrac{dx}{ds} \dfrac{ds}{dt} = \dfrac{\beta_l  x'}{1 + x/\rho} = \dfrac{\beta p_x}{1 + \Delta p/p} \label{betar}\\
\beta_y &=& \dfrac{1}{c}\dfrac{dy}{dt} = \dfrac{1}{c}\dfrac{dy}{ds} \dfrac{ds}{dt} = \dfrac{\beta_l  y'}{1 + x/\rho} = \dfrac{\beta p_y}{1 + \Delta p/p} \label{betay}\\
\beta_l &=& \left(\beta^2 - \beta_r^2 - \beta_y^2\right)^{1/2} = \dfrac{\beta}{1 + \Delta p/p} \left[ (1 + \Delta p/p)^2 - p_x^2 - p_y^2 \right]^{1/2}
\end{eqnarray}

\subsection{Calculation of the spin precession vectors}
Injecting Eqs. (\ref{betar}) and (\ref{betay}) into the expression of $\Omega_l$ and $\Omega_r$, one obtains:
\begin{eqnarray}
\Omega_l &=& \dfrac{e}{mc} \left[a+\dfrac{1}{\gamma+1} \right](\beta_r E_y-\beta_y E_r) \nonumber \\
         &=& \dfrac{e}{mc} \left[a+\dfrac{1}{\gamma+1} \right] \dfrac{\beta_l}{1 + x/\rho} (x' E_y-y' E_r)
\end{eqnarray}
and 
\begin{eqnarray}
\Omega_r &=& \dfrac{e}{mc} \left[a+\dfrac{1}{\gamma+1} \right](\beta_y E_l-\beta_l E_y) \nonumber \\
         &=& \dfrac{e}{mc} \left[a+\dfrac{1}{\gamma+1} \right] \beta_l \left(\dfrac{y'}{1 + x/\rho} E_l-E_y \right)
\end{eqnarray}
By means of Eq. (\ref{rho}), the horizontal spin precession vector can be calculated for an electrostatic bend:
\begin{eqnarray}
\Omega_y &=&  \dfrac{e}{mc} \left[a+\dfrac{1}{\gamma+1} \right] ( \beta_l E_r - \beta_r E_l) +  \dfrac{1}{\rho} \dfrac{ds}{dt} \nonumber \\
&\approx & \dfrac{e}{mc} \beta_l E_r^b \left[a + \dfrac{1}{\gamma +1} - \dfrac{1}{\gamma_m \beta_m ^2} \right] + \dfrac{e}{mc} \left[a + \dfrac{1}{\gamma + 1} \right] \beta_l (E_r -E_r^b) - \beta_l c \dfrac{x}{\rho ^2} \nonumber \\
& & -\dfrac{e}{mc} \left[a+\dfrac{1}{\gamma+1} \right] \dfrac{\beta_l}{1 + x/\rho} x' E_l \label{omegay}
\end{eqnarray}
where $E_r^b$ is the design radial electric field of the electrostatic deflector. \\
If $\gamma = \gamma_m$, i.e. for a particle at the reference energy and with no transverse displacements ($x=x'=y=y'=0$), 
\begin{eqnarray}
a + \dfrac{1}{\gamma + 1} - \dfrac{1}{\gamma_m \beta_m^2} &=& a- \dfrac{1}{\beta_m^2 \gamma_m^2} \nonumber \\
&=& a -\left(\dfrac{mc}{p} \right)^2
\end{eqnarray}
The above quantity vanishes at the momentum $p = mc /\sqrt{a}$ which is generally referred to as the magic momentum since then, this guarantees that $\Omega_r = \Omega_y = \Omega_l =0$, i.e. the spin precession vanishes. This is referred to as the frozen spin condition and is considered as a robust way to improve the sensitivity of the EDM measurement for charged particles. The main idea is to keep the spin aligned along the momentum vector of the particle and to provide a strong enough radial electric field acting on the EDM for the duration of the beam storage time \cite{farley}.  \\ 
For protons, this is achieved for,
\begin{eqnarray}
p_m = \dfrac{mc}{\sqrt{a}} = \dfrac{938.2721 \mbox{\text{ MeV/c}}}{\sqrt{1.79285}} = 700.7398 \text{ MeV/c} \label{magic_momentum}
\end{eqnarray}
corresponding to a proton kinetic energy of $E_m \sim 232.79$ MeV. \\
In reality though, the particle momentum deviates from the magic value. Thus, it is important to evaluate the impact of any momentum offset on the spin precession rate. \\
Back to Eq. (\ref{omegay}). Let's assume that $\gamma \neq \gamma_m$. Then, one can write:
\begin{eqnarray}
a + \dfrac{1}{\gamma + 1} - \dfrac{1}{\gamma_m \beta_m^2} &=& -\dfrac{1}{\gamma_m + 1} + \dfrac{1}{\gamma + 1} \hspace{2mm};\hspace{2mm} a = \dfrac{1}{\beta_m^2 \gamma_m^2} \nonumber \\
&=& -\dfrac{1}{\gamma_m + 1} + \dfrac{1}{(\gamma_m + 1)}  \dfrac{1}{\left[1 + (\gamma - \gamma_m)/(\gamma_m + 1) \right]} \nonumber \\
&= & -\dfrac{\gamma - \gamma_m}{(\gamma_m + 1)^2} + \dfrac{(\gamma - \gamma_m)^2}{(\gamma_m + 1)^3} - \dfrac{(\gamma - \gamma_m)^3}{(\gamma_m + 1)^4} + ...
\end{eqnarray}
Substituting Eq. (\ref{gamma}) into the first order term of the above expression yields:
\begin{eqnarray}
\gamma_m - \gamma &=& \left[1 + \left(\dfrac{p_m c}{E_0} \right)^2 \right]^{1/2} - \left[1 + \left(\dfrac{p_m c}{E_0}\right)^2 \left(1 +\dfrac{\Delta p}{p} \right)^2 \right]^{1/2} \nonumber \\
&\approx & \dfrac{-\left(\dfrac{p_m c}{E_0} \right)^2}{\left[1+\left(\dfrac{p_m c}{E_0} \right)^2 \right]^{1/2}} \left[ \dfrac{\Delta p}{p} + \dfrac{1}{2 {\gamma_m}^2} \left(\dfrac{\Delta p}{p} \right)^2 \right] 
\nonumber \\
& = & \dfrac{-1}{a \left(1 + \dfrac{1}{a}\right)^{1/2}} \dfrac{\Delta p}{p} - \dfrac{1}{2 a \left(1 + \dfrac{1}{a}\right)^{3/2}} \left(\dfrac{\Delta p}{p} \right)^2
\end{eqnarray}
Injecting the latter into the expression of $\Omega_y$, one finally obtains:
\begin{eqnarray}
\Omega_y &\approx & \dfrac{-e}{mc} \beta_l E_r^b \left[\dfrac{1}{a (\gamma_m +1)^2 \gamma_m} \dfrac{\Delta p}{p} + \dfrac{2\gamma_m - a(\gamma_m +1)}{2 (\gamma_m +1)^3 \gamma_m ^3 a^2} \left(\dfrac{\Delta p}{p} \right)^2 \right] + \dfrac{e}{mc} \left[a + \dfrac{1}{\gamma + 1} \right] \beta_l (E_r -E_r^b) \nonumber \\
& & - \beta_l c \dfrac{x}{\rho ^2} - \dfrac{e}{mc} \left[a+\dfrac{1}{\gamma+1} \right] \dfrac{\beta_l}{1 + x/\rho} x' E_l
\end{eqnarray}
In the above derivation, it is important to note that the spin precession vector is calculated with respect to the momentum of the reference particle orbit. In summary, the expressions of the spin precession components are as follows:
\begin{equation}
\begin{cases}
\Omega_y = K - \dfrac{e}{mc} \left[a+\dfrac{1}{\gamma+1} \right] \dfrac{\beta_l}{1 + x/\rho} x' E_l \\ \\
\Omega_l = \dfrac{e}{mc} \left[a+\dfrac{1}{\gamma+1} \right] \dfrac{\beta_l}{1 + x/\rho} (x' E_y-y' E_r) \\ \\
\Omega_r = \dfrac{e}{mc} \left[a+\dfrac{1}{\gamma+1} \right] \beta_l \left(\dfrac{y'}{1 + x/\rho} E_l-E_y \right) \\ \\
K \approx \left\{
  \begin{array}{@{}ll@{}}
    \dfrac{-e/mc}{ a \left(\gamma_m +1 \right)^2 \gamma_m} \beta_l E_r^b \dfrac{\Delta p}{p} + \dfrac{e}{mc} \left[a + \dfrac{1}{\gamma + 1} \right] \beta_l (E_r -E_r^b)  - \beta_l c \dfrac{x}{\rho ^2} & \text{if bend} \\ \\
    \dfrac{e}{mc} \left[a+\dfrac{1}{\gamma+1} \right] \beta_l E_r & \text{otherwise}
  \end{array}\right.
\end{cases}
\label{quant}
\end{equation}

\section{On solving the Thomas BMT equation using perturbation method}
Eq. (\ref{matrix}) can also be written in the matrix form:
\begin{eqnarray}
\dfrac{d\bm{S}}{dt} = \bm{M}(t) \bm{S}(t) = \epsilon \bm{M}_1(t)\bm{S}(t) = \epsilon \bm{Y}(t,\bm{S}) \label{spineq}
\end{eqnarray}
where $\epsilon > 0$ is a small parameter. Thus, the Bogoliubov-Krylov-Mitropolski (BKM) method of averages can be invoked \cite{BKM}. The main idea of the averaging method is based on the assumption that the derivative $d\bm{S}/dt$ is small, which is typically the case for the frozen spin method considered here (see (Eq. \ref{magic_momentum})). Thus, the evolution of $\bm{S}$ contains two contributions (obeying two timescales): a slowly varying term $\bm{\xi}$ due to the presence of $\epsilon$ in front of $\bm{M}_1$, and small rapidly oscillating terms, due to the presence of $t$ in $\bm{M}_1$, i.e. describing the spin precession changes within the elements. \\
The basic idea of this approach was first developed by Krylov and Bogoliubov (1934). Later on, in 1958, Bogoliubov and Mitropolski established the general scheme and a more rigorous treatment for this method. Finally, in 1969, Perko almost completed the theory with error estimates for the periodic and quasi-periodic cases \cite{perko}. \\
It is assumed that the vector function $\bm{Y}(t,\bm{S})$ possesses an average value:
\begin{eqnarray}
\mean{\bm{Y}(t,\bm{S})}= \lim_{T\to\infty} \dfrac{1}{T} \int_{0}^{T} \bm{Y}(t,\bm{S}) dt \label{averages}
\end{eqnarray}
This condition is satisfied for all cases considered here given that $\bm{Y}$ is almost periodic in time. The asymptotic solution of Eq. (\ref{spineq}) can be obtained in the following way: in a first approximation in terms of $\epsilon$, one can write:
\begin{eqnarray}
\bm{S}(t) = \bm{\xi}(t)+ \epsilon \bm{\tilde{Y}}(t,\bm{\xi}) \label{asym}
\end{eqnarray}
where $\bm{\xi}(t)$ satisfies the equation:
\begin{eqnarray}
\dfrac{d\bm{\xi}}{dt} = \epsilon \mean{\bm{Y}(t,\bm{\xi})} = \mean{\bm{M}(t)} \bm{\xi} \label{13}
\end{eqnarray}
The average is to be taken only with respect to the explicit variable $t$ and the tilde $\bm{\sim}$ is the integrating operator defined by:
\begin{eqnarray}
\bm{\tilde{Y}}(t,\bm{\xi}) = \int \left[\bm{Y}(t,\bm{\xi}) - \mean{\bm{Y}(t,\bm{\xi})} \right] dt \label{tilde}
\end{eqnarray}
From Eq. (\ref{spineq}), (\ref{asym}) and (\ref{tilde}), it results that a first order approximate solution to the T-BMT equation where the frozen spin condition is quasi-satisfied ($\epsilon$ small) is given by:
\begin{eqnarray}
\bm{S}(t) = \left[\bm{1} + \bm{\tilde{M}}(t) \right] \bm{\xi}(t) \label{1orderbogoo}
\end{eqnarray}

To simplify the analysis, let's solve for the slowly varying term in $\bm{S}$. From what preceded (Eq. \ref{13}), the zero order approximation of $\bm{S}$ is a solution of:
\begin{eqnarray}
\dfrac{d\bm{\xi}}{dt} =\mean{\bm{M}(t)} \bm{\xi}
\end{eqnarray}
In order to apply the above formalism, one has to compute the average of the matrix $\bm{M}(t)$ by tracking the particle over a sufficient number of turns, or in a simpler way by searching for the closed orbit which satisfies the periodicity of $\bm{M}(t)$ and thus the averages are defined over one full turn. \\ \\
NB: In the above definitions, $\mean{}$, $\bm{\sim}$ represent operators acting on all the elements of the matrix.  \\ \\
$\mean{M}$ is a skew symmetric matrix. So all eigenvalues are imaginary. 
Solving for the eigenvalues $\lambda_i$, one obtains:
\begin{eqnarray}
\lambda_1 &=& 0 \\
\lambda_2 &=& -i \sqrt{\mean{\Omega_r}^2+\mean{\Omega_y}^2+\mean{\Omega_l}^2} = -i \mean{\Omega}\\
\lambda_3 &=& i \sqrt{\mean{\Omega_r}^2+\mean{\Omega_y}^2+\mean{\Omega_l}^2} = i \mean{\Omega}
\end{eqnarray}
and the corresponding eigenvectors are:
\[
\renewcommand\arraystretch{1.5}
\bm{v_{\lambda_1}} = 
    \begin{pmatrix}
        \frac{\mean{\Omega _r}}{\mean{\Omega _l}}  \\
        \frac{\mean{\Omega _y}}{\mean{\Omega _l}} \\
        1
    \end{pmatrix} 
\hspace{2mm};\hspace{2mm}
\renewcommand\arraystretch{1.5}
\bm{v_{\lambda_2}} = 
    \begin{pmatrix}
\frac{\mean{\Omega _r} \mean{\Omega _y} + i \mean{\Omega _l} \mean{\Omega}}{\mean{\Omega _l} \mean{\Omega _y}- i \mean{\Omega _r} \mean{\Omega}} \\
-\frac{\mean{\Omega _l}^2+\mean{\Omega _r}^2}{\mean{\Omega _l} \mean{\Omega _y}-i\mean{\Omega _r} \mean{\Omega}} \\
        1
    \end{pmatrix} 
\hspace{2mm};\hspace{2mm}
\renewcommand\arraystretch{1.5}
\bm{v_{\lambda_3}} = 
    \begin{pmatrix}
    \frac{\mean{\Omega _r} \mean{\Omega _y} - i \mean{\Omega _l} \mean{\Omega}}{\mean{\Omega _l} \mean{\Omega _y}+ i \mean{\Omega _r} \mean{\Omega}} \\
-\frac{\mean{\Omega _l}^2+\mean{\Omega _r}^2}{\mean{\Omega _l} \mean{\Omega _y}+i\mean{\Omega _r} \mean{\Omega}} \\
        1
    \end{pmatrix}    
\]
A general solution to this equation writes in the form: $\bm{\xi}(t)= c_1 \bm{v_{\lambda_1}} e^{\lambda_1 t} + c_2 \bm{v_{\lambda_2}} e^{\lambda_2 t} + c_3 \bm{v_{\lambda_3}} e^{\lambda_3 t}$.
Now, solving for \textbf{$\xi$} such that \textbf{$\xi$(0)}=(0,0,1), one obtains the zero order solution:
\begin{equation}
\begin{cases}
\xi_r(t)=\frac{\mean{\Omega_l}  \mean{\Omega_r} }{\mean{\Omega}^2} \left[1- \cos \left(t \mean{\Omega} \right) \right] + \frac{\mean{\Omega_y}}{\mean{\Omega}}   \left[\sin \left(t \mean{\Omega} \right) \right] \\ \\
\xi_y(t)= \frac{\mean{\Omega_l}  \mean{\Omega_y}}{\mean{\Omega}^2} \left[1-\cos \left(t \mean{\Omega}\right) \right]-\frac{\mean{\Omega_r}}{\mean{\Omega}} \left[\sin \left(t \mean{\Omega} \right) \right] \\ \\
\xi_l(t)= 1+\frac{\mean{\Omega_y}^2+\mean{\Omega_r} ^2}{\mean{\Omega}^2} \left[\cos \left(t \mean{\Omega} \right)-1 \right] \label{xi}
\end{cases}
\end{equation}
Assuming that the argument of the $\cos$ and $\sin$ functions is small (EDM signal level), then Eq. (\ref{xi}) may be rewritten:
\begin{eqnarray}
\xi_r(t) &\approx & \dfrac{\mean{\Omega_l} \mean{\Omega_r}}{2} t^2 + \mean{\Omega_y} t  \\
\xi_y(t) &\approx & \dfrac{\mean{\Omega_l} \mean{\Omega_y}}{2} t^2 - \mean{\Omega_r} t \\
\xi_l(t) &\approx & 1 - \dfrac{\mean{\Omega_y}^2 + \mean{\Omega_r}^2}{2} t^2 \label{approx_frozen}
\end{eqnarray}
To verify the validity of the above solutions, one can compute the norm of $\bm{\xi}(t)$ and it can be shown that \mbox{for any given $t$,} $\parallel\bm{\xi}(t)\parallel=({\xi_r}^2+{\xi_y}^2+{\xi_l}^2)^{1/2}=\parallel\bm{\xi}(0)\parallel=1$. The conservation of the Hermiticity of the zero order solutions is thus verified.
From Eqs. (\ref{xi}) and (\ref{1orderbogoo}) it results that, to the first order in the method of averages, the solution of the T-BMT equation is given by:
\begin{eqnarray}
\begin{cases}
s_r(t) = \xi_r(t) + \tilde{\Omega}_y(t) \xi_l(t) - \tilde{\Omega}_l(t) \xi_y(t)  \\ \\
s_y(t) = \xi_y(t) - \tilde{\Omega}_r(t) \xi_l(t) + \tilde{\Omega}_l(t) \xi_r(t)  \\ \\
s_l(t) = \xi_l(t) - \tilde{\Omega}_y(t) \xi_r(t) + \tilde{\Omega}_r(t) \xi_y(t)\label{s1storder}
\end{cases}
\end{eqnarray}
where
\begin{eqnarray}
\tilde{\Omega}_i(t) = \int_0^t \left[\Omega_i(\tau) - \mean{\Omega_i} \right] d\tau \hspace{2mm};\hspace{2mm} i = r,y,l
\end{eqnarray}
In the above derivation, it is assumed that $\bm{S}$ changes slowly. This is formulated in such a way that the right hand side term of Eq. (\ref{spineq}) is proportional to $\epsilon$, therefore $\bm{S} \sim \epsilon$. Thus, $\bm{S}$ changes appreciably only in a large time interval $t \sim 1/\epsilon$. Of course, the above argument is not rigorous. However, the foundation of the method of averaging was established by  N. N. Bogoliubov who demonstrated that one can construct a system of first approximation and higher orders, the solutions of which approximate the solutions of the initial system within an arbitrary degree of accuracy. It is important to note that the tilde parameters in Eq. (\ref{s1storder}) give rise only to small vibrations around $\xi$ and do not influence the systematic change of the spin. \\
Now, let's characterize the evolution of the tilde parameters on the closed orbit:

\subsubsection*{Property}
\textit{``On the closed orbit, the tilde parameters vanish after the completion of each turn.''}\\ \\
To establish this, let's consider a quantity $x$. $x$ refers to the field or the coordinates of the particle. Such a quantity defined on the closed orbit is periodic in $T$. Thus, without any loss of generality, $x$ can be written in the following way:
\begin{eqnarray}
x(t) = \mean{x} + g_x(t)
\end{eqnarray} 
where $g_x(t)$ is a periodic function in $T$ such that $\mean{g_x}=0$.
Now,
\begin{eqnarray}
\tilde{x}(t) = \int_{0}^{t} \left[x(\tau)-\mean{x} \right] d\tau = \int_{0}^{t} g_x(\tau) d\tau
\end{eqnarray}
Thus, given that $\mean{g_x}=0$, one obtains
\begin{eqnarray}
\tilde{x}(nT) = n \int_{0}^{T} g_x(\tau) d\tau = 0
\end{eqnarray}

\subsection{Spin rotation matrix}
We start from
\begin{eqnarray}
\dfrac{d\xi}{dt} = \mean{M(t)} \xi(t)
\end{eqnarray}
Then, assuming that $\xi(t) = R(t) \xi(t_0)$ where $R(t)$ is the spin rotation matrix satisfying $R(t_0)=I$, one obtains:
\begin{eqnarray}
\dfrac{dR}{dt} = \mean{M(t)} R(t) \hspace{2mm};\hspace{2mm} R(t_0)=I
\end{eqnarray}
The solution is obtained using the Euler-Rodrigues formula:
\begin{eqnarray}
R(t) = e^{\mean{M}t}= I + \mean{M} \dfrac{\sin(\mean{\Omega}t)}{\mean{\Omega}} + \mean{M}^2 \dfrac{1-\cos(\mean{\Omega}t)}{\mean{\Omega}^2}
\end{eqnarray}
Now, assuming that $\xi(0)=(0,0,1)$, one obtains:
\begin{align}
\renewcommand\arraystretch{1.8}
    \begin{pmatrix}
        \xi_r \\
        \xi_y \\
        \xi_l
    \end{pmatrix}_{CW}
&=
\begin{pmatrix}
\frac{\mean{\Omega _y} \sin \left(t \mean{\Omega} \right)}{\mean{\Omega}}+\frac{\mean{\Omega _l} \mean{\Omega _r} \left(1-\cos \left(t \mean{\Omega} \right)\right)}{\mean{\Omega}^2} \\
\frac{-\mean{\Omega _r} \sin \left(t \mean{\Omega} \right)}{\mean{\Omega}}+\frac{\mean{\Omega _l} \mean{\Omega _y} \left(1-\cos \left(t \mean{\Omega} \right)\right)}{\mean{\Omega}^2} \\
\frac{\left(-\mean{\Omega _r}^2-\mean{\Omega _y}^2\right) \left(1-\cos \left(t \mean{\Omega} \right)\right)}{\mean{\Omega}^2}+1
\end{pmatrix}
\approx
\begin{pmatrix}
\mean{\Omega_y} t + \dfrac{\mean{\Omega_l} \mean{\Omega_r}}{2} t^2 \\
-\mean{\Omega_r} t + \dfrac{\mean{\Omega_l} \mean{\Omega_y}}{2} t^2  \\
 1 - \dfrac{\mean{\Omega_y}^2 + \mean{\Omega_r}^2}{2} t^2
\end{pmatrix}
\end{align}
which is the same result as obtained in Eq. (\ref{xi}) above.

\subsection{2nd order approximation}
In reality, when solving for the vertical spin component, the buildup is determined by the product of the rapidly oscillating terms of the particle motion (velocities and experienced fields) with the rapidly oscillating spin components. In certain configurations where the average value of the rapidly oscillating terms of the spin precession vanishes \mbox{i.e. $\mean{\Omega}=0$,} a spin buildup still exists. This is due to the non-commutativity of spin rotations around different axes and is generally referred to as geometric phase. Such a phenomena, also known as Pancharatnam-Berry phase \cite{geom1} or often Berry phase \cite{geom2} is defined as the phase difference acquired over the course of a cycle, when a system is subjected to cyclic adiabatic processes. Therefore, to account for such a contribution to the vertical spin buildup, a second order approximation is employed in which one can integrate the Thomas BMT equation by re-injecting the first order solution based on the method of averages. First, one re-writes the radial spin buildup from the exact T-BMT equation:
\begin{eqnarray}
\dfrac{ds_r}{dt} = \Omega_y s_l - \Omega_l s_y \label{dsrdtn}
\end{eqnarray}
where $s_l$ and $s_y$ are solved using the method of averages (Eq. \ref{s1storder}):
\begin{eqnarray}
s_l(t) &=& \xi_l(t) + \phi_l(t) \nonumber \\
s_y(t) &=& \xi_y(t) + \phi_y(t)
\end{eqnarray}
and $\phi_l$, $\phi_y$ are the rapidly oscillating terms that vanish after each turn completion (on the closed orbit) as established in the property above and defined by:
\begin{eqnarray}
\phi_l(t) &=& - \tilde{\Omega}_y(t) \xi_r(t) + \tilde{\Omega}_r(t) \xi_y(t) \\
\phi_y(t) &=& -\tilde{\Omega}_r(t) \xi_l(t) + \tilde{\Omega}_l(t) \xi_r(t)
\end{eqnarray}
Next, one injects this improved first approximation into the exact equation (\ref{dsrdtn}). This yields:
\begin{eqnarray}
\dfrac{ds_r}{dt} &=& \left[\Omega_y(t) \xi_l(t) - \Omega_l(t) \xi_y(t) \right] + \left[ \Omega_y(t) \left(- \tilde{\Omega}_y(t) \xi_r(t) + \tilde{\Omega}_r(t) \xi_y(t) \right) + \Omega_l(t) \left( \tilde{\Omega}_r(t) \xi_l(t) - \tilde{\Omega}_l(t) \xi_r(t) \right) \right] \\
&=& \left[\Omega_y(t) + \Omega_l(t) \tilde{\Omega}_r(t) \right] \xi_l(t) + \left[-\Omega_y(t)\tilde{\Omega}_y(t) - \Omega_l(t) \tilde{\Omega}_l(t) \right] \xi_r(t) + \left[-\Omega_l(t) + \Omega_y(t) \tilde{\Omega}_r(t) \right] \xi_y(t)
\end{eqnarray}
Assuming that $\xi_y(t), \xi_r(t) \ll \xi_l(t)$, and keeping the terms up to the second order, one can integrate the above equation (assuming $\xi_l \approx 1 $ and $\xi_y \approx -\mean{\Omega_r} t$) to compute the radial spin component:
\begin{eqnarray}
s_r(t) \approx \mean{\Omega_y(t) + \Omega_l(t) \tilde{\Omega}_r(t)} t + \tilde{\Omega}_y(t) + \widetilde{\Omega_l(t) \tilde{\Omega}_r(t)} + \mean{\Omega_r}  \int_0 ^t d\tau \tau \Omega_l(\tau)  \label{sr_geom0}
\end{eqnarray}
Now, one expresses $\Omega_l$ in the following way:
\begin{eqnarray}
\Omega_l(t) &=& \left( \Omega_l(t) - \mean{\Omega_l} \right) + \mean{\Omega_l} \nonumber \\
&=& \dfrac{d}{dt} \tilde{\Omega}_l + \mean{\Omega_l}
\end{eqnarray}
Thus, by means of an integration per parts, the expression of the integral can be simplified:
\begin{eqnarray}
\int_0^t d\tau \Omega_l(\tau) \tau &=& \int_0^t d\tau \mean{\Omega_l} \tau + \int_0^t d\tau \dfrac{d}{d\tau} \tilde{\Omega}_l(\tau) \tau \nonumber \\
&=& \dfrac{\mean{\Omega_l}}{2} t^2 + \left[\tau \tilde{\Omega}_l \right]_0^t - \int_0^t d\tau \tilde{\Omega}_l \nonumber \\
&=&  \dfrac{\mean{\Omega_l}}{2} t^2 + t\tilde{\Omega}_l(t) - \mean{\tilde{\Omega}_l} t - \widetilde{\tilde{\Omega}}_l(t) \label{tomegal}
\end{eqnarray}
so that Eq. (\ref{sr_geom0}) re-writes in the following way:
\begin{eqnarray}
s_{r,2}(t) = \xi_{r,2}(t) + \phi_{r,2}(t)
\end{eqnarray}
where the subscript 2 denotes the second order approximation for which the frozen solution is given by:
\begin{eqnarray}
\boxed{\xi_{r,2}(t) = \mean{\Omega_y} t + \mean{\Omega_l \tilde{\Omega}_r} t - \mean{\Omega_r} \mean{\tilde{\Omega}_l} t + \dfrac{\mean{\Omega_r}\mean{\Omega_l}}{2}t^2}
\end{eqnarray}
and the rapidly oscillating terms that vanish after each turn completion are:
\begin{eqnarray}
\boxed{\phi_{r,2}(t) = \tilde{\Omega}_y(t) + \widetilde{\Omega_l(t) \tilde{\Omega}_r(t)} + \mean{\Omega_r} \left[t\tilde{\Omega}_l(t) - \widetilde{\tilde{\Omega}}_l(t) \right]}
\end{eqnarray}
Thus, the second order approximation yielded additional terms for the frozen solution in comparison with the \mbox{$1^{st}$ order} approximation:
\begin{itemize}
\item The first additional term $\mean{\Omega_l \tilde{\Omega}_r} t$ accounts for the geometric phases induced by the rapid oscillations of the vertical spin component (although its average is neglected) and longitudinal spin precession. As will be established later on, such a contribution becomes dominant at the proximity to the magic energy ($\mean{\Omega_y} \approx 0$).
\item The second additional term $-\mean{\Omega_r} \mean{\tilde{\Omega}_l} t$ is due to the slowly linear varying term of the vertical spin component coupled with the rapidly oscillating terms of the longitudinal spin precession.
\end{itemize}
Next, one repeats the same steps to compute the vertical spin buildup from the exact T-BMT equation:
\begin{eqnarray}
\dfrac{ds_y}{dt} = -\Omega_r(t) s_l(t) + \Omega_l(t) s_r(t) \label{dsydtn}
\end{eqnarray}
where $s_r$ and $s_l$ are solved using the method of averages (Eq. \ref{s1storder}):
\begin{eqnarray}
s_r(t) &=& \xi_r(t) + \phi_r(t) \nonumber \\
s_l(t) &=& \xi_l(t) + \phi_l(t)
\end{eqnarray}
and $\phi_r$, $\phi_l$ are the rapidly oscillating terms that vanish after each turn completion (on the closed orbit) as established in the property above and defined by:
\begin{eqnarray}
\phi_r(t) &=& \tilde{\Omega}_y(t) \xi_l(t) + \tilde{\Omega}_l(t) \xi_y(t) \\
\phi_l(t) &=& - \tilde{\Omega}_y(t) \xi_r(t) + \tilde{\Omega}_r(t) \xi_y(t)
\end{eqnarray}
Next, one injects this improved first approximation into the exact equation (\ref{dsydtn}). This yields:
\begin{eqnarray}
\dfrac{ds_y}{dt} &=& \left[ -\Omega_r(t) \xi_l(t) + \Omega_l(t) \xi_r(t) \right] + \left[ \Omega_r(t) \left( \tilde{\Omega}_y(t) \xi_r(t) - \tilde{\Omega}_r(t) \xi_y(t) \right)  + \Omega_l(t) \left(\tilde{\Omega}_y(t) \xi_l(t) - \tilde{\Omega}_l(t) \xi_y(t) \right) \right] \\
&=& \left[ -\Omega_r(t)+\Omega_l(t) \tilde{\Omega}_y(t) \right] \xi_l(t) + \left[\Omega_l(t)+\Omega_r(t) \tilde{\Omega}_y(t) \right] \xi_r(t) + \left[-\Omega_r(t) \tilde{\Omega}_r(t) -\Omega_l(t) \tilde{\Omega}_l(t) \right] \xi_y(t)
\end{eqnarray}
where the third (Right Hand Side) term can be neglected. 
Now, injecting the approximate expressions \mbox{of $\xi$ ($\xi_l \approx 1$, $\xi_r \approx \mean{\Omega_y} t$)}, keeping the terms up to the second order and integrating over time, one obtains:
\begin{eqnarray}
s_y(t) \approx \mean{-\Omega_r(t)+\Omega_l(t) \tilde{\Omega}_y(t)} t - \tilde{\Omega}_r(t) + \widetilde{\Omega_l(t) \tilde{\Omega}_y(t)} + \mean{\Omega_y(t)} \int_0 ^t d\tau \tau \Omega_l(\tau) \label{precise_sy}
\end{eqnarray}
and injecting Eq. (\ref{tomegal}) into the above expression, one can write a second order approximation to the vertical spin component:
\begin{eqnarray}
s_{y,2}(t) = \xi_{y,2}(t) + \phi_{y,2}(t)
\end{eqnarray}
where the subscript 2 denotes the second order approximation for which the frozen solution is given by:
\begin{eqnarray}
\boxed{ \xi_{y,2}(t) \approx -\mean{\Omega_r} t + \mean{\Omega_l \tilde{\Omega}_y} t - \mean{\Omega_y} \mean{\tilde{\Omega}_l} t + \dfrac{\mean{\Omega_y} \mean{\Omega_l}}{2} t^2 } \label{precise_sysimpl}
\end{eqnarray}
and the rapidly oscillating terms that vanish after each turn completion are:
\begin{eqnarray}
\boxed{\phi_{y,2}(t)= -\tilde{\Omega}_r(t) + \widetilde{\Omega_l(t) \tilde{\Omega}_y(t)} + \mean{\Omega_y} \left[t\tilde{\Omega}_l(t) - \widetilde{\tilde{\Omega}}_l(t) \right]} \label{eq:phiy}
\end{eqnarray}
In summary, the second order frozen solution yields 4 terms contributing to the vertical spin buildup:
\begin{enumerate}
\item The first contribution is due to the longitudinal spin component coupled with the average contribution of the radial spin precession. This accounts for any net average radial magnetic field for instance.
\item The second contribution is due to the rapidly oscillating terms of the radial spin components coupled with the longitudinal spin precession. This accounts for the geometric phases as well for any net average longitudinal spin precession.
\item The third contribution is due to the slowly linearly varying term of the radial spin component coupled with the rapidly oscillating terms of the longitudinal spin precession.
\item The fourth and last contribution is due to the slowly linearly varying term of the radial spin component coupled with the average contribution of the longitudinal spin precession (quadratic increase).
\end{enumerate}

Finally, one repeats the same steps to compute the second order approximation of the longitudinal spin buildup:
\begin{eqnarray}
\dfrac{ds_l}{dt} &=& -\Omega_y(t) s_r(t) + \Omega_r(t) s_y(t) \nonumber \\
&=& \left[-\Omega_y(t) \tilde{\Omega}_y(t) - \Omega_r(t) \tilde{\Omega}_r(t) \right] \xi_l(t) + \left[-\Omega_y(t) + \Omega_r(t) \tilde{\Omega}_l(t) \right] \xi_r(t) + \left[\Omega_r(t) + \Omega_y(t) \tilde{\Omega}_l(t) \right] \xi_y(t)
\end{eqnarray}
Assuming $\xi_l \approx 1$, $\xi_r \approx \mean{\Omega_y} t$, $\xi_y \approx -\mean{\Omega_r} t$, one can integrate the above equation. First, one can establish the following identity:
\begin{eqnarray}
\int_0 ^t d\tau \Omega_r(\tau) \tilde{\Omega}_r(\tau) = \dfrac{[{\tilde{\Omega}_r}(t)]^2}{2} + \mean{\Omega_r}\mean{\tilde{\Omega}_r}t + \mean{\Omega_r} \widetilde{\tilde{\Omega}}_r(t)
\end{eqnarray}
so that one can solve for the longitudinal spin component which writes in the following way:
\begin{eqnarray}
\begin{cases}
s_{l,2}(t) = \xi_{l,2}(t) + \phi_{l,2}(t) \\ \\
\xi_{l,2}(t) = 1- \dfrac{\mean{\Omega_y}^2+\mean{\Omega_r}^2}{2}t^2 \\ \\
\phi_{l,2}(t) = -\mean{\Omega_r} t \tilde{\Omega}_r - \mean{\Omega_y} t \tilde{\Omega}_y - \dfrac{[{\tilde{\Omega}_r}(t)]^2 + [{\tilde{\Omega}_y}(t)]^2}{2}
\end{cases}
\end{eqnarray}
\textbf{Remarks}: 
\begin{itemize}
\item The second order approximation above is modified from the original BKM method of averages: if one follows the original method, the second order approximation is obtained from the exact T-BMT equation by re-injecting the first order approximation and taking the averages with respect to the time as an explicit variable while considering $\bm{\xi}$ constant. Such an assumption is not rigorous since it neglects the impact of the product of the linearly varying terms of $\bm{\xi}$ with the rapid oscillations of the spin precession components $\bm{\Omega}$.
\item If the average value of $\Omega_l$ or $\Omega_y$ is zero, then the quadratic evolution with time of the vertical spin vanishes. Therefore, clearing the frozen radial spin component by means of a feedback system can eliminate the quadratic buildup. However, this is only valid up to the second order. Higher order terms yield additional quadratic time dependence.
\item The frozen solution $\xi$ (see Eq. (\ref{precise_sysimpl})) is only valid at time $t$ which is an integer number of the revolution period. In other terms if the spin components are recorded after each turn completion. Therefore it is a natural quantity to describe the signal measured at the location of the polarimeter.
\item If $\mean{\Omega_r} = \mean{\Omega_y} = \mean{\Omega_l} = 0$ then $\xi_{y,2}(t) \approx \mean{\Omega_l \tilde{\Omega}_y} t $ which accounts for the geometric phases to the second order.
\item The above calculation can be easily extended to the next order. Although there seems to be no need in general to pursue this, one might be interested though to determine the high order terms of the Berry phases. The final result of the third order frozen spin solution is given by:
\begin{eqnarray}
\begin{cases}
\xi_{y,3}(t) = A_{1} t + A_{2} t^2 + A_{3} t^3 \\ \\
A_{1} = -\mean{\Omega_r} \left[1+ \mean{\Omega_l \widetilde{\tilde{\Omega}}_l} + \mean{\widetilde{\Omega_r \tilde{\Omega}}_r} + \mean{\tilde{\Omega}_l}^2 + \mean{\Omega_l}\mean{\tilde{\Omega}_l}-\mean{\Omega_l}\mean{\widetilde{\tilde{\Omega}}_l} - \mean{\Omega_r} \mean{\widetilde{\tilde{\Omega}}_r}  \right] + \mean{\Omega_l \tilde{\Omega}_y} + \mean{\Omega_l \widetilde{\Omega_l \tilde{\Omega}_r}} \nonumber \\
\hspace{10mm} - \mean{\Omega_y} \left[ \mean{\tilde{\Omega}_l} + \mean{\widetilde{\Omega_r \tilde{\Omega}}_y} - \mean{\Omega_y} \mean{\widetilde{\tilde{\Omega}}_r} \right] + \dfrac{\mean{\Omega_r \left[(\tilde{\Omega}_r)^2+(\tilde{\Omega}_y)^2 \right]}}{2} - \mean{\Omega_l \tilde{\Omega}_r} \mean{\tilde{\Omega}_l} \\ \\
A_{2} = \dfrac{\mean{\Omega_y} \mean{\Omega_l}}{2} + \dfrac{\mean{\Omega_y}}{2} \left[\mean{\Omega_r \tilde{\Omega}_y} - \mean{\Omega_y} \mean{\tilde{\Omega}_r} \right] + \dfrac{\mean{\Omega_l}}{2} \left[\mean{\Omega_l \tilde{\Omega}_r} - \mean{\Omega_r}\mean{\tilde{\Omega}_l} \right] \\ \\
A_3 = \mean{\Omega_r} \dfrac{\mean{\Omega_r}^2+\mean{\Omega_y}^2}{6}+\dfrac{\mean{\Omega_r}\mean{\Omega_l}^2}{6}
\end{cases}
\end{eqnarray}
It results that the only contribution of the geometric phases up to the third order is given by:
\begin{eqnarray}
\boxed{\xi_{y,3}(t) = \mean{\Omega_l \tilde{\Omega}_y} t + \mean{\Omega_l \widetilde{\Omega_l \tilde{\Omega}_r}} t - \mean{\Omega_l \tilde{\Omega}_r} \mean{\tilde{\Omega}_l}t + \dfrac{\mean{\Omega_r (\tilde{\Omega}_y)^2}}{2} t} \label{Eq:geom_phases}
\end{eqnarray}
since $\mean{\Omega_r (\tilde{\Omega}_r)^2} = 0$ if $\mean{\Omega_r}=0$.
\end{itemize}
This will be discussed later on in this paper.

\subsection{Error analysis}
The above second (and third) order approximation to the T-BMT equation is based on the assumption that the average spin precession component is negligible on the timescales of the EDM experiment. If the spin coherence time is $T_{coh} = 1000 \text{ s}$ as is generally assumed to reach the sensitivity required to measure the EDM, then the previous assumption can be formulated as follows:
\begin{eqnarray}
\mean{\Omega} T_{coh} = \sqrt{\mean{\Omega_r}^2 + \mean{\Omega_y}^2 + \mean{\Omega_l}^2} T_{coh} \ll 1 \hspace{2mm} \Rightarrow \hspace{2mm} \mean{\Omega_r}, \mean{\Omega_y}, \mean{\Omega_l} \ll \dfrac{1}{T_{coh}} \approx 10^{-3}
\end{eqnarray}
From the above scheme one can infer that the general frozen solution to the T-BMT equation can be classified into three main regimes depending on the value of the average spin precession:
\begin{enumerate}
\item If $0<\mean{\Omega} \ll 1/T_{coh}$ then any non-linear increase with time of the spin can be neglected on the timescales of the EDM experiment. And using the $N^{th}$ order approximation based on the Bogoliubov method of averages, it can be seen that:
\begin{eqnarray}
\xi_y(t) = \xi_{y,n}(t) + \mathcal{O}\left(\mean{\Omega}^{n+1} t\right) 
\end{eqnarray} 
For instance, the second order approximation satisfies the T-BMT equation to an accuracy of the order \mbox{of $\mean{\Omega}^3 t$.} \\
In general though, the non-linearities arising from the solution can be neglected for short timescales in comparison with $T_{coh}$ ($\sim$ milliseconds).
\item If $\mean{\Omega} \gtrsim 1/T_{coh}$ then the spin evolution is mainly governed by the averages of the spin precession components if the same property holds for all components i.e. $\mean{\Omega_y},\mean{\Omega_r},\mean{\Omega_l} \gtrsim 1/T_{coh}$. Therefore, the first order solution (Eq. (\ref{xi})) is the most accurate one.
\item In the limit where $\mean{\Omega}=0$, i.e. $\mean{\Omega_r}=\mean{\Omega_l}=\mean{\Omega_y}=0$, the geometric phases are the only contribution to the spin buildup. The latter is governed by the non-commutativity of the rotation around different axes. There appears no obvious way to estimate the error of the solution in this particular case, since the spin buildup is mainly determined by how the errors are distributed in the ring with respect to each other. However the third order approximation (see Eq. (\ref{Eq:geom_phases})) contains the dominant contributions. Note that the geometric phases which arise initially as a linear function of time can be transferred from one plane to the other therefore yielding additional terms. 
\end{enumerate}
It is important to note that the Hermiticity of the approximate frozen solution is only conserved for the $1^{st}$ order terms of the approximation. To show this, let's consider the second order approximation in the particular case where $\mean{\Omega}=0$ and calculate the norm of the frozen solution only at times $t=kT$ i.e. after each turn completion:
\begin{eqnarray}
\parallel\bm{\xi}(t=kT)\parallel=({\xi_{l,2}}^2+{\xi_{r,2}}^2+{\xi_{y,2}}^2)^{1/2} &=& \left( 1 + \left[\mean{\Omega_l \tilde{\Omega}_r}^2 + \mean{\Omega_l \tilde{\Omega}_y}^2 \right] t^2 \right)^{1/2} \nonumber \\
&\approx & 1 + \dfrac{\mean{\Omega_l \tilde{\Omega}_r}^2 + \mean{\Omega_l \tilde{\Omega}_y}^2}{2} t^2
\end{eqnarray} 
Such an effect is negligible for the timescales of the EDM experiment. However, the Hermiticity can be improved by keeping the higher order terms in the expansion of the sinusoidal functions of the $1^{st}$ order solution. This will not be pursued here since the EDM experiment requires slowly varying parameters of the average spin precession.

\subsection{Benchmarking of the analytical solution with numerical simulations}
To establish the validity of the analytical solutions, one simulated several cases by solving the T-BMT differential equations using explicit Runge Kutta tracker in MATHEMATICA. The spin precession components are defined as a linear combination of sinusoidal functions. For all cases considered, one assumed $\mean{\Omega_r}=0$ to simplify the analysis. For more details regarding the numerical solution, see Appendix \ref{app:A}. Note that, for all cases considered here, unless otherwise specified, one relies on the second order approximation since the first order approximation is particularly inaccurate when the averages of the spin precession components vanish. This is generally the case at the proximity to the magic energy.
\subsubsection{Geometric phases}
\begin{figure*}%
    \centering
    \subfloat[Radial spin versus time.]{{\includegraphics[width=8cm]{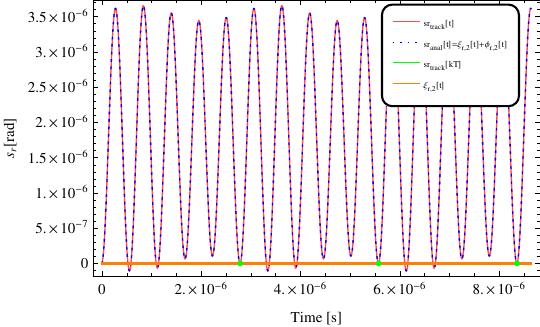} }}%
    \qquad
    \subfloat[Vertical spin versus time.]{{\includegraphics[width=8cm]{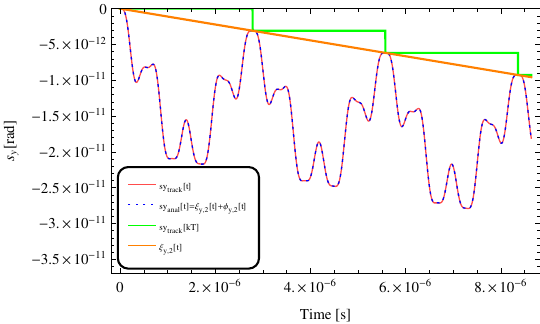} }}%
    \caption{Comparison of the analytical estimate with the tracking simulations in MATHEMATICA. The tracking results are shown in red while the analytical ones are shown in blue dashed lines. The stepwise function (in green) displays the tracking results evaluated after each turn completion which coincide with the frozen solution (orange).}%
    \label{fig:geom_ph}%
\end{figure*}

\begin{figure}
\centering 
\includegraphics*[width=10cm]{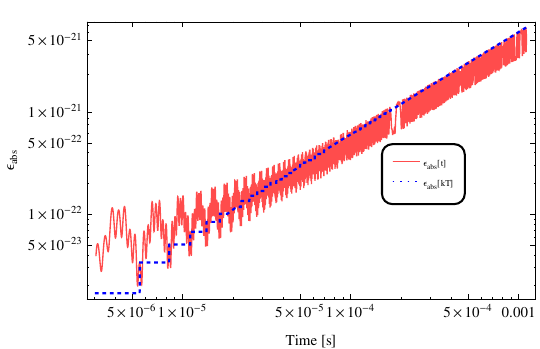}
\caption{Logarithmic plot of the absolute error versus time of the vertical spin component. It can be seen from the plot that the absolute error increases linearly with time so that the relative error between the analytical and the numerical solution remains constant: $\epsilon_{rel}=5*10^{-10} \%$.}
\label{fig:geome_ph_error}
\end{figure}

In the first case considered, one assumes $\mean{\Omega}=0$. The result is that the vertical spin buildup is determined by the geometric phases such as \mbox{$\xi_{r,2}(t) \approx \mean{\Omega_l \tilde{\Omega}_r} \approx 0$ and $\xi_{y,2}(t) \approx \mean{\Omega_l \tilde{\Omega}_y} t = -1.1*10^{-6} t $ rad}. Such an effect is always present and is strongly related to how the errors are distributed in the ring. A comparison of the numerical simulation with the analytical result is shown in \mbox{fig \ref{fig:geom_ph}.}
We define $\epsilon_{abs}$ (resp. $\epsilon_{rel}$) as the absolute error (resp. relative error) between the convergent numerical solution and the analytical one. Fig \ref{fig:geome_ph_error} shows the absolute error which grows linearly with time (the relative error is therefore constant).

\subsubsection{Geometric phases and growing radial spin component}
\begin{figure}
\centering 
\includegraphics*[width=10cm]{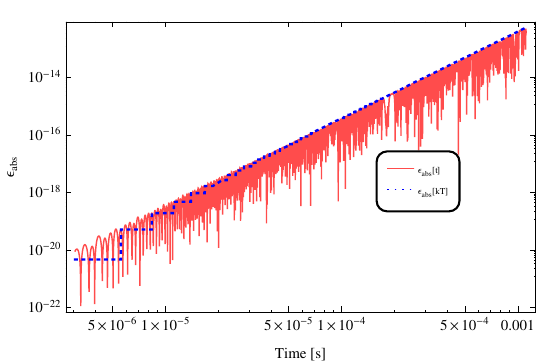}
\caption{Logarithmic plot of the absolute error versus time of the vertical spin component. In this case, the absolute error can be approximated by $\epsilon_{abs}(t) \approx 4*10^{-4}*t^3$. }
\label{fig:geome_ph2_error}
\end{figure}

In this case, $\mean{\Omega}=\mean{\Omega_y}=10$ $\text{rad/s}$ which typically occurs when the particle is not at the magic energy. This yields a non-vanishing frozen radial spin component \mbox{$\xi_{r,2}(t) \approx \mean{\Omega_y}t=-10*t$ rad} as well as an additional term contributing to the vertical spin buildup such as \mbox{$\xi_{y,2}(t) \approx \mean{\Omega_l \tilde{\Omega}_y} t - \mean{\Omega_y} \mean{\tilde{\Omega}_l} t = -1.1*10^{-6} t - 2.2*10^{-5} t $ rad}.
\begin{figure*}%
    \centering
    \subfloat[Radial spin versus time.]{{\includegraphics[width=8cm]{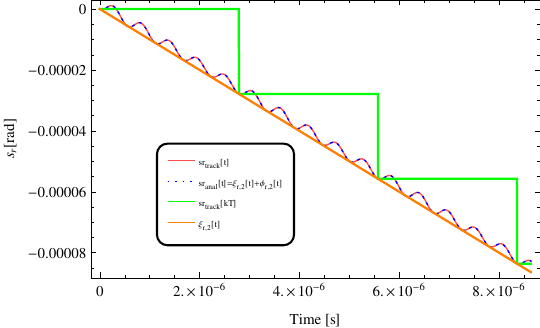} }}%
    \qquad
    \subfloat[Vertical spin versus time.]{{\includegraphics[width=8cm]{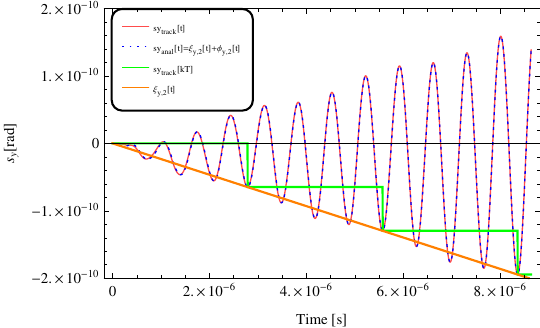} }}%
    \caption{Comparison of the analytical estimate with the tracking simulations in MATHEMATICA.}%
    \label{fig:geom_ph2}%
\end{figure*}
A comparison of the analytical results with the numerical simulations is shown in fig \ref{fig:geom_ph2} where one can observe a good agreement. However, as illustrated in fig \ref{fig:geome_ph2_error}, the error grows cubically with time. This is explained by the fact that in the above derivations, one only considered the first order linear term in the expansion of $\xi_r$ (Eq. \ref{xi}). This is valid for very small timescales. However, as the time grows, the approximation may no longer be valid, and this is particularly enhanced here by the fact that \mbox{$\mean{\Omega}=10$ rad/s} is non negligible. Thus, one needs to include the next higher order term in the expansion, which yields the cubical time dependence since $\sin(t)\approx t -t^3/6$. \\
Note as well that the vertical spin oscillates with an amplitude growing over time. As predicted by \mbox{Eq. (\ref{eq:phiy}),} this is due to $\mean{\Omega_y} \neq 0$. Thus, it is advantageous to use more than one polarimeter located at specific azimuthal locations around the ring to probe such imperfections.

\subsubsection{Geometric phases, growing radial spin component and quadratic increase}
\begin{figure*}%
    \centering
    \subfloat[Radial spin versus time.]{{\includegraphics[width=8cm]{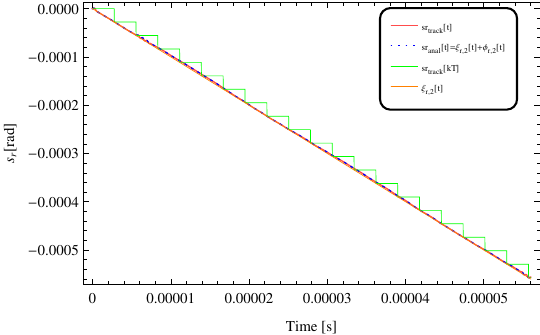} }}%
    \qquad
    \subfloat[Vertical spin versus time.]{{\includegraphics[width=8cm]{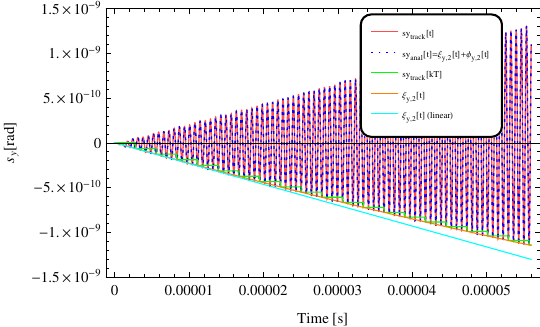} }}%
    \caption{Comparison of the analytical estimate with the tracking simulations in MATHEMATICA.}%
    \label{fig:geom_ph3}%
\end{figure*}

\begin{figure}
\centering 
\includegraphics*[width=10cm]{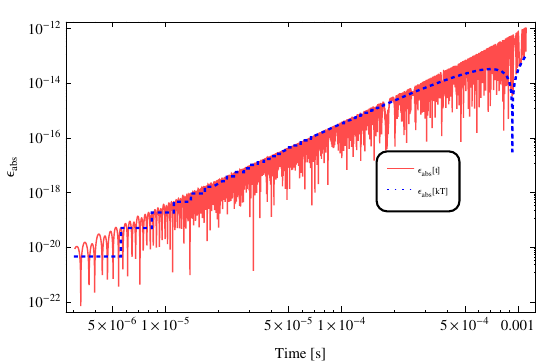}
\caption{Logarithmic plot of the absolute error versus time of the vertical spin component. In this case, the absolute error can be approximated by $\epsilon_{abs}(t) \approx 4*10^{-4}*t^3-0.4*t^4$. }
\label{fig:geome_ph3_error}
\end{figure}

In this case, one added a non-vanishing longitudinal spin precession such as $\mean{\Omega_l}=10^{-2}$ rad/s in comparison with the previous case. This yields an additional quadratic spin buildup so that: 
\begin{eqnarray}
\xi_{y,2}(t) &\approx & \mean{\Omega_l \tilde{\Omega}_y} t - \mean{\Omega_y} \mean{\tilde{\Omega}_l} t + \dfrac{\mean{\Omega_y} \mean{\Omega_l}}{2} t^2  \\
&=& -1.1*10^{-6} t - 2.2*10^{-5} t  + 0.05 * t^2 \mbox{rad}
\end{eqnarray}
Figure \ref{fig:geom_ph3} displays the tracking results as well as a comparison with the analytical formula for the first 20 revolution periods where one can observe the impact of the quadratic spin buildup in comparison with the linear one. The error (see fig \ref{fig:geome_ph3_error}) is similar to the previous case although one can observe the existence of additional higher order terms.  

\subsubsection{No vertical spin precession and zero averages}
\begin{figure*}%
    \centering
    \subfloat[Radial spin versus time.]{{\includegraphics[width=8cm]{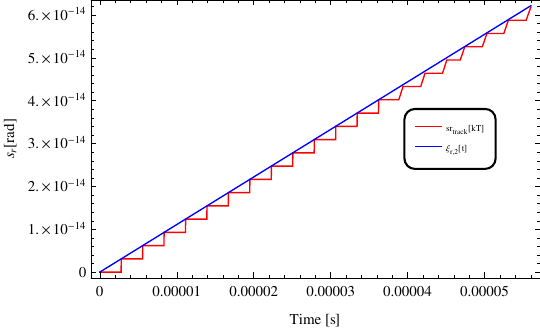} }}%
    \qquad
    \subfloat[Vertical spin versus time.]{{\includegraphics[width=8cm]{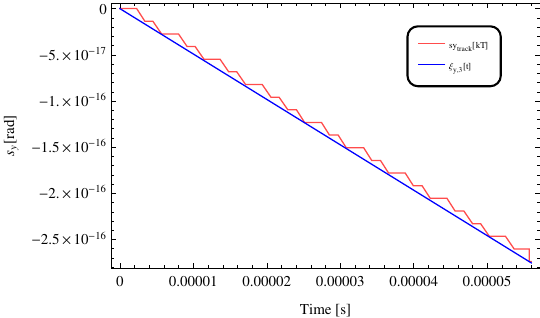} }}%
    \caption{Comparison of the analytical estimate with the tracking simulations in MATHEMATICA. The radial spin buildup is due to the geometric phases. The latter are transferred from the horizontal to the vertical plane following consecutive rotations around the longitudinal axis.}%
    \label{fig:geom_ph4}%
\end{figure*}

A question that arises frequently is whether a beam injected at the magic energy so that the vertical spin precession vanishes everywhere can still yield a vertical spin buildup? To answer this, one assumes no vertical spin precession as well as zero averages of the spin precession components in the ring, \mbox{i.e. $\Omega_y(t)=0$ and $\mean{\Omega_l}=\mean{\Omega_r}=0$.} It results from the second order approximation that:
\begin{eqnarray}
s_{r,2}(t) = \mean{\Omega_l \tilde{\Omega}_r}t + \widetilde{\Omega_l(t) \tilde{\Omega}_r(t)} \label{sr_geom_zero}
\end{eqnarray}
so that the radial spin increase results from the net effect of the spin rotations around the longitudinal and radial axes.
However, given that $\Omega_y(t)=0$, the second order approximation yielded no vertical spin buildup which is not in agreement with the numerical simulations. This is a sign that higher order terms are missing. To derive the 3rd order approximation of the vertical spin component in this particular case, one re-injects Eq. (\ref{sr_geom_zero}) into (\ref{dsydtn}). This yields:
\begin{eqnarray}
\dfrac{ds_y}{dt} = -\Omega_r(t) + \Omega_l(t) \left[ \mean{\Omega_l \tilde{\Omega}_r}t + \widetilde{\Omega_l(t) \tilde{\Omega}_r(t)} \right]
\end{eqnarray}
or, after integration:
\begin{eqnarray}
s_y(t) = -\tilde{\Omega}_r(t) - \mean{\Omega_l \tilde{\Omega}_r} \mean{\tilde{\Omega}_l}t + \mean{\Omega_l \tilde{\Omega}_r} \left[t \tilde{\Omega}_l(t) - \widetilde{\tilde{\Omega}}_l(t) \right] + \mean{\Omega_l \widetilde{\Omega_l \tilde{\Omega}_r}} t + \widetilde{\Omega_l(t) \widetilde{\Omega_l \tilde{\Omega}_r}}
\end{eqnarray}
so that the frozen spin solution is approximated by:
\begin{eqnarray}
\begin{cases}
\xi_{r,2}(t) = \mean{\Omega_l \tilde{\Omega}_r}t = 1.1*10^{-9} \text{$t$ rad}\\
\xi_{y,3}(t) = -\mean{\Omega_l \tilde{\Omega}_r} \mean{\tilde{\Omega}_l}t + \mean{\Omega_l \widetilde{\Omega_l \tilde{\Omega}_r}} t = 2.45*10^{-15} \text{$t$} -4.9*10^{-12} \text{$t$ rad}
\end{cases}
\end{eqnarray}
Thus, in the above system, the vertical spin buildup arises first by means of radial spin precession. This leads to a linear radial spin buildup via rotations around the longitudinal axis that do not commute with the radial one. Such a radial spin buildup acts back on the vertical one by means of rotations around the longitudinal axis. These rotations transfer the geometric phases from the horizontal plane to the vertical one. The following diagram summarizes this scheme:
\[
s_l \xrightarrow[\text{}]{\mathbf{\Omega_r}} s_y=\phi_y=-\tilde{\Omega}_r(t) \xrightarrow[\text{}]{\mathbf{\Omega_l}} s_r=\xi_{r,2}+\phi_{r,2} \xrightarrow[\text{}]{\mathbf{\Omega_l}} s_y=\xi_{y,3}+\phi_{y,3} 
\]
and comparison of the tracking simulations with the analytical results are shown in fig \ref{fig:geom_ph4}.
Therefore, in a fully electrostatic ring, and for a beam injected at the magic energy, vertical spin buildup can occur by means of consecutive rotations around the longitudinal and radial axes induced by the electric fields and the non-vanishing slope of the particle trajectories. \\
Furthermore, assuming $\Omega_l(t)=0$ and $\mean{\Omega_r}=\mean{\Omega_y}=0$, it can be verified (although it will not be discussed here) that the vertical spin buildup can occur due to the rapidly oscillating terms of the longitudinal spin component coupled with the radial spin precession such as \mbox{$s_y(t) \approx 1/2 \mean{\Omega_r(\tilde{\Omega}_y)^2} t$}.

\subsection{System oscillating with several frequencies}
In reality, the spin precession components can be oscillating with different frequencies, not necessarily equal to the particle revolution frequency. In this case, the question arises of whether there exists an efficient way to compute the averages of the previous equations without having to track a particle for a consequent number of turns? The question can also be formulated differently: can we define a period $T$ associated to the system under which the spin oscillations are represented as a sum of slowly varying terms and rapidly rotating terms that vanish at times $t=kT, k$ integer? 
To simplify the analysis, let's assume that the system is oscillating with two frequencies: the revolution frequency of the particle since the stable motion is always near the closed orbit, and a second frequency perturbing the spin motion. The latter can be due to any external or internal force of the beam, mismatching issue, etc. For instance, the vertical spin precession component can be written in the following way:
\begin{eqnarray}
\Omega_y(t)= f_1\left(\dfrac{2\pi}{T_{rev}}t\right)+f_2\left(\dfrac{2\pi}{T_{per}}t\right)
\end{eqnarray}
where $f_1$ and $f_2$ are two periodic functions in $t$ with periods $T_{rev}$ and $T_{per}$ respectively.
Now, assuming there exists a rational number $p/q$ ($p,q$ mutually prime numbers) such as:
\begin{eqnarray}
\dfrac{T_{rev}}{T_{per}} &=& \dfrac{p}{q} \\
\Rightarrow q T_{rev} &=& p T_{per} = T
\end{eqnarray}
this implies that $T$ is the frequency of the system as defined above. In reality, $T_{rev}/T_{per}$ is a real number. Therefore, the idea is to obtain its closest rational approximation. This can be done with the help of continued fractions.

\subsection{Accuracy of the numerical simulations}
In general, the higher the order of the integration method, the more accurate is the numerical scheme. Accuracy here refers to the smallness of the local error, i.e. the error introduced after each integration step of the differential equation. Therefore, to improve the accuracy of the solution, it is generally assumed that smaller step size is a sufficient condition. However, when stiff systems are considered, some of the solutions obtained can be highly unstable. Besides, it appears that the accuracy of the obtained solutions is highly sensitive to the step size chosen.
Stiffness is mainly encountered when hard edge models are assumed, i.e. due to an abrupt change in the particle field or its coordinates. In addition, the total energy conservation is a fundamental criterion to assess the accuracy of the numerical simulations as is discussed in \cite{selcuk}. Using the explicit Runge Kutta tracker available in MATHEMATICA, one investigated the impact of the stiffness of the T-BMT equaton on the error of the calculation (see appendix \ref{app:B}). 

\clearpage 

\section{Spin tracking simulations in an all-electric ring lattice}
\begin{figure}
\centering 
\includegraphics*[width=10cm]{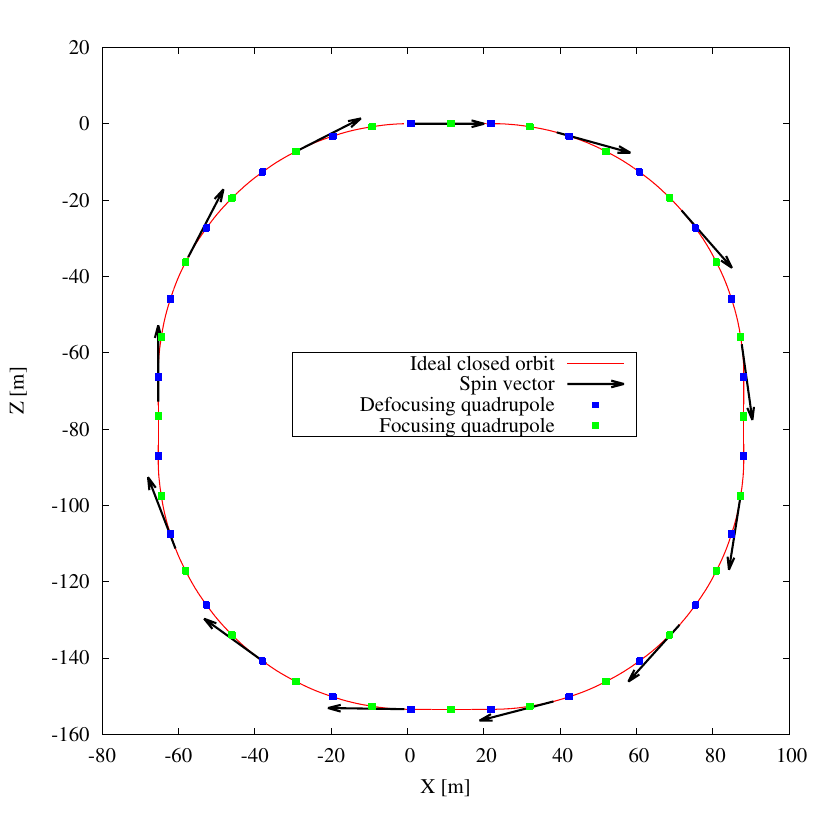}
\caption{Overview of the ideal ring. The beam is circulating clockwise such as the particle momentum vector is aligned with the spin vector.}
\label{lattice}
\end{figure}

As established earlier, when the proton is at the magic energy (Eq. (\ref{magic_momentum})), the spin is frozen in the horizontal plane along the momentum direction. Such a condition is met at one specific energy for a purely electric ring. \\
In what follows one simulated the all-electric Lebedev ring lattice in BMAD based on the storage ring EDM collaboration proposal \cite{lebedev}\cite{sredm}. The ring structure consists of 4 superperiods, each including 5 FODO cells with 3 cylindrical deflectors per half cell. An overview of the simulated ring is shown in fig \ref{lattice} in the ideal case where the particle momentum vector is perfectly aligned with the spin vector. The main ring and beam parameters are summarized in table \ref{param_EDM}. 
\renewcommand{\arraystretch}{1.5}
\begin{table}[!th]
\centering
\begin{tabular}{|*1{p{50mm}}|*1{p{36mm}}|}
\cline{1-2}
 Total beam energy & 1.171 GeV\\
\cline{1-2}
Focusing structure  & FODO\\
\cline{1-2}
  $N_{cells}$, number of cells & 20\\
\cline{1-2}
Deflector shape & cylindrical \\
\cline{1-2}
Number of deflectors per cell & 6 \\
\cline{1-2}
Bending radius $\rho$ & 52.3089 m \\
\cline{1-2}
 Radial E field  & 8.016 MV/m\\
\cline{1-2}
Gap & 3 cm \\
\cline{1-2}
Bending voltage & $\pm$ 120 kV \\
\cline{1-2}
Horizontal tune $Q_x$ & 2.42 \\
\cline{1-2}
Vertical tune $Q_y$ & 0.44 \\
\cline{1-2}
\end{tabular}
\vspace{2mm}
\caption{Table of the ring parameters of the proton EDM experiment.}
\label{param_EDM}
\end{table}

\begin{figure}
\centering 
\includegraphics*[width=12cm]{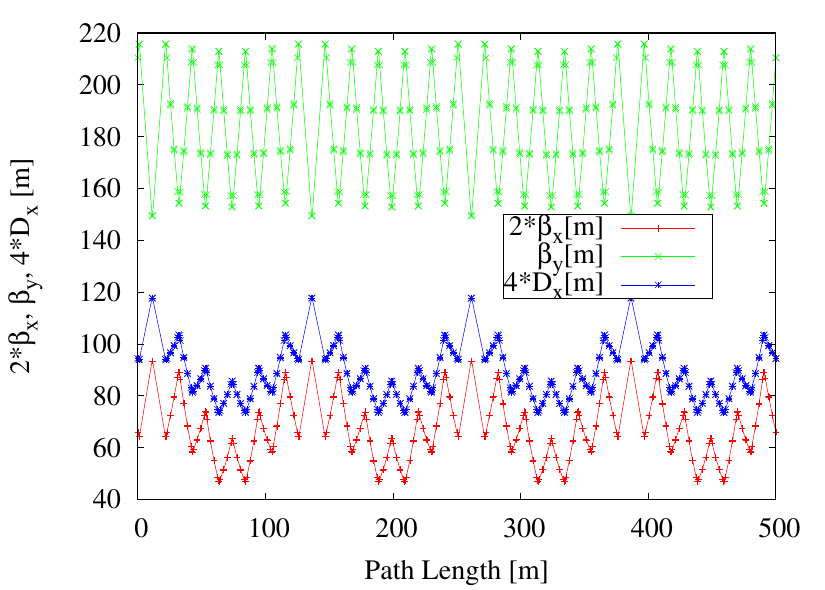}
\caption{Twiss parameters for the entire circumference of the all-electric ring lattice.}
\label{twiss}
\end{figure}

Fig \ref{twiss} shows the Twiss parameters in the ring. In particular, the weak vertical focusing, $\beta_y^{max}=216$ m, was chosen in order to achieve enough sensitivity on the radial magnetic field.
The aim of the remaining sections of this paper is to benchmark the previously established analytical formula against tracking simulations. The approach relies on computing the averages of the spin precession components on the closed orbit (no betatron motion considered) and comparing the results with the tracking simulations.

\clearpage

\section{Case with no longitudinal and no magnetic field components}
In order to undertake the benchmarking analysis, one needs to reduce the complexity of the problem. For that reason, one first considers a lattice which is free of magnetic fields as well as any longitudinal electric field imperfections $E_l$ = 0. Thus, the spin precession components may be written:
\begin{equation} 
\begin{cases}
\Omega_y = K \\ \\
\Omega_l = \dfrac{e}{mc} \left[a+\dfrac{1}{\gamma+1} \right] \dfrac{\beta_l}{1 + x/\rho} (x' E_y-y' E_r) \\ \\
\Omega_r = -\dfrac{e}{mc} \left[a+\dfrac{1}{\gamma+1} \right] \beta_l E_y \\ \\
K = \left\{
  \begin{array}{@{}ll@{}}
    \dfrac{-e/mc}{ a \left(\gamma_m +1 \right)^2 \gamma_m} \beta_l E_r^b \dfrac{\Delta p}{p} + \dfrac{e}{mc} \left[a + \dfrac{1}{\gamma + 1} \right] \beta_l (E_r -E_r^b)  - \beta_l c \dfrac{x}{\rho ^2} & \text{if bend} \\ \\
    \dfrac{e}{mc} \left[a+\dfrac{1}{\gamma+1} \right] \beta_l E_r & \text{otherwise}
  \end{array}\right.
\end{cases}
\label{average_quant}
\end{equation}
From the first order approximation (Eq. (\ref{s1storder})), one can write for very short timescales:
\begin{eqnarray}
s_r(t) &=& \xi_r(t) + \tilde{\Omega}_y(t) \xi_l(t) - \tilde{\Omega}_l(t) \xi_y(t) \approx \xi_r(t) + \tilde{\Omega}_y(t) \label{srad} \\
s_y(t) &=& \xi_y(t) - \tilde{\Omega}_r(t) \xi_l(t) + \tilde{\Omega}_l(t) \xi_r(t) \approx \xi_y(t) - \tilde{\Omega}_r(t) \label{svert}
\end{eqnarray}
so that one can compute the rapidly oscillating terms for the vertical spin component:
\begin{eqnarray}
\tilde{\Omega}_r(t) = \dfrac{-e}{mc} \left[a + \dfrac{1}{\gamma + 1} \right] \beta_l \int_{0}^{t} \left[E_y(\tau) - \mean{E_y} \right] d\tau  \label{sycom1}
\end{eqnarray}
Now, for a fully eletrostatic ring, the particle equation of motion in the vertical plane may be written:
\begin{eqnarray}
e E_y = \dfrac{dp_y}{dt}
\end{eqnarray}
which yields after integration
\begin{eqnarray}
\int_{0}^{t} E_y(\tau) d\tau = \dfrac{1}{e} \left[p_y(t) - p_y(0)\right] = \dfrac{\beta_l \gamma mc}{e} \left[y'(t)-y'(0) \right] \label{yp}
\end{eqnarray}
Thus, on the closed orbit, $\mean{E_y} = 0$ and consequently $\mean{\Omega_r}=0$. Injecting Eq. (\ref{yp}) into (\ref{sycom1}) yields:
\begin{eqnarray}
\phi_y(t) &\approx & \left[y'(t) - y'(0)\right] \label{sycom2}
\end{eqnarray}
To the first order, the rapidly oscillating terms of the vertical spin components are mainly determined by the vertical particle slope.

\subsection{Benchmarking of the analytical solution with numerical simulations for very small timescales}
\begin{figure}
\centering 
\includegraphics*[width=12cm]{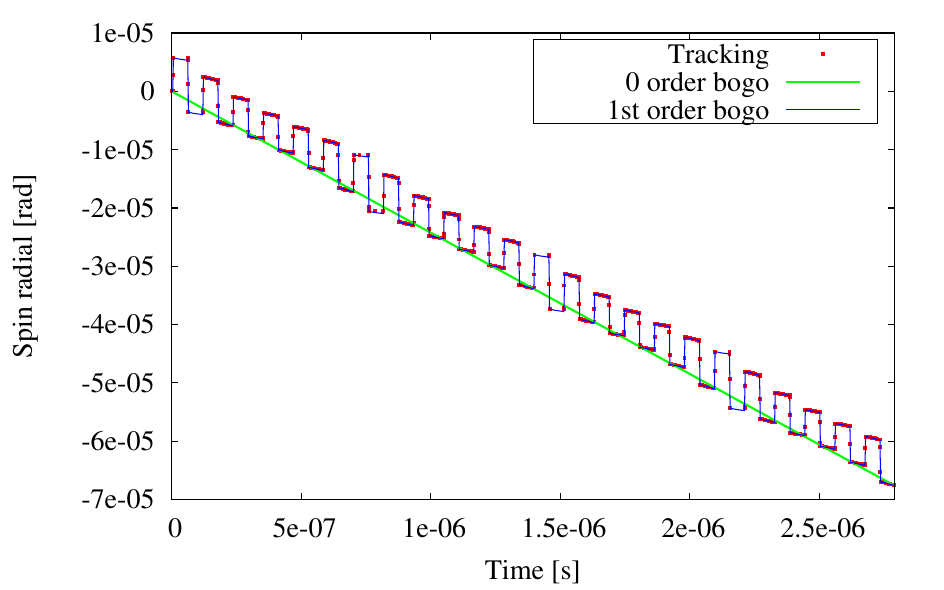}
\caption{Comparison of the approximate solution of the radial spin component with the tracking results for one revolution period.}
\label{WKB_sr}
\end{figure}

\begin{figure}
\centering 
\includegraphics*[width=12cm]{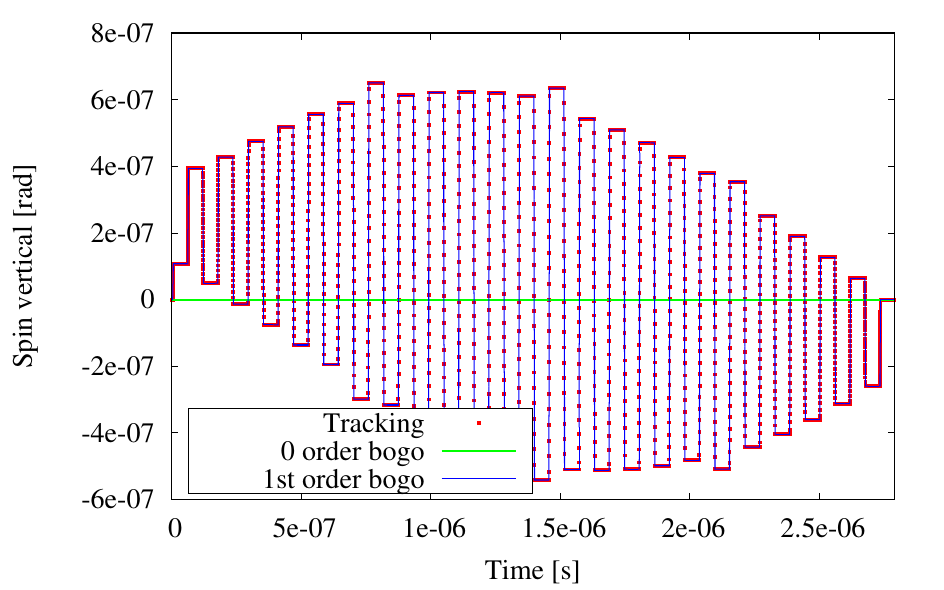}
\caption{Comparison of the approximate solution of the vertical spin component with the tracking results for one revolution period.}
\label{WKB_sy}
\end{figure}

Now, let's compare the analytical solution with the results of tracking simulations in a fully electrostatic ring in the case where $t \to 0$, i.e. for a few turns from injection: a particle is launched with an initial momentum offset $\Delta p/p = 10^{-5}$ and tracked over several 1000 of turns to compute its trajectory. One defocusing quadrupole is vertically misaligned in the lattice by $\Delta y = 10 \text{ $\mu$m}$ thereby yielding a vertical orbit motion. The spin is computed by solving the Thomas BMT equation in the Frenet Serret coordinate system using a Runge Kutta integrator available in BMAD. From the orbit tracking results, one can compute the averages as well as the tilde of all quantities needed in \mbox{Eq. (\ref{average_quant}).} \\ \\
Comparison of the tracking results with Eqs. (\ref{srad}) and (\ref{sycom2}) is shown in figures \ref{WKB_sr} and \ref{WKB_sy} respectively. The agreement is quite good for both quantities which demonstrates that the perturbation approach based on the method of averages is an adequate way to obtain an approximate solution of the T-BMT equation since all quantities evolve adiabatically in time. In particular, it is shown that the rapid oscillating terms  for the vertical spin component are determined by the vertical angles of the particle. From Eq. (\ref{sycom2}), it follows that a first order approximation for very short timescales is given by:
\begin{eqnarray}
s_y(t) \approx y'(t)-y'(0) \hspace{2mm}\mbox{for}\hspace{2mm} t \to 0 
\end{eqnarray}
Note that the above analysis still holds if the particle is off the closed orbit. However, in that case, the averages should be calculated over several turns in order to remove the fluctuations of the spin precession components.

\subsection{Study of misalignment errors}
Next, one compares the previously established analytical formula with the tracking results without restriction to very short timescales. The Lebedev lattice \cite{lebedev} is modified by introducing a vertical misalignment of one or several quadrupoles. The spin is only recorded after each turn completion. Thus, the benchmarking analysis with the analytical formula is based on the frozen spin solution. We discuss two cases, depending on the value of the average spin precession. Some of the following simulation results were discussed in previous publications \cite{selcuk_quad}.
\subsubsection{Case of a beam injected off the magic momentum}
\begin{figure}
\centering 
\includegraphics*[width=12cm]{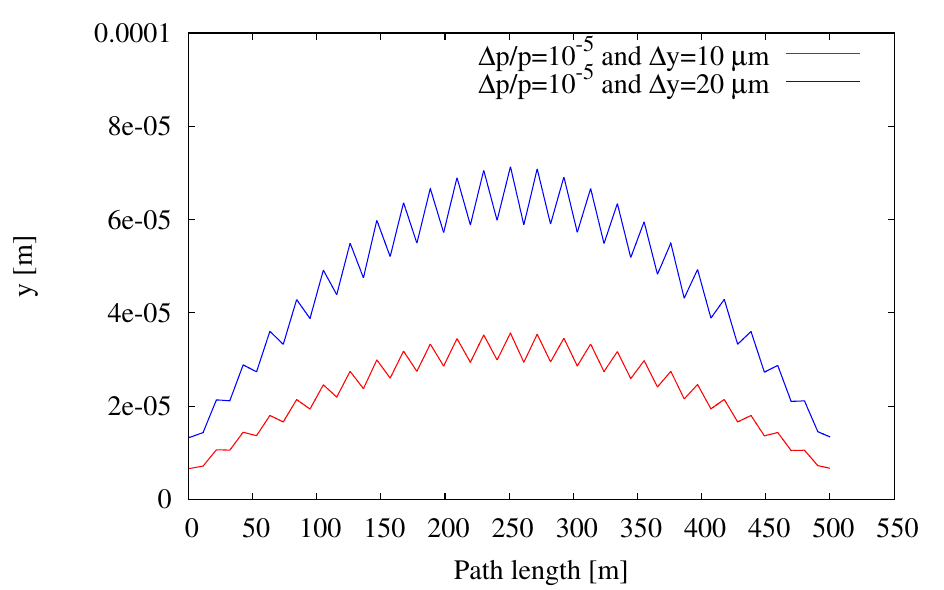}
\caption{Vertical closed orbit as a function of the path length.}
\label{vert_CO}
\end{figure}

\begin{figure}
\centering 
\includegraphics*[width=10cm]{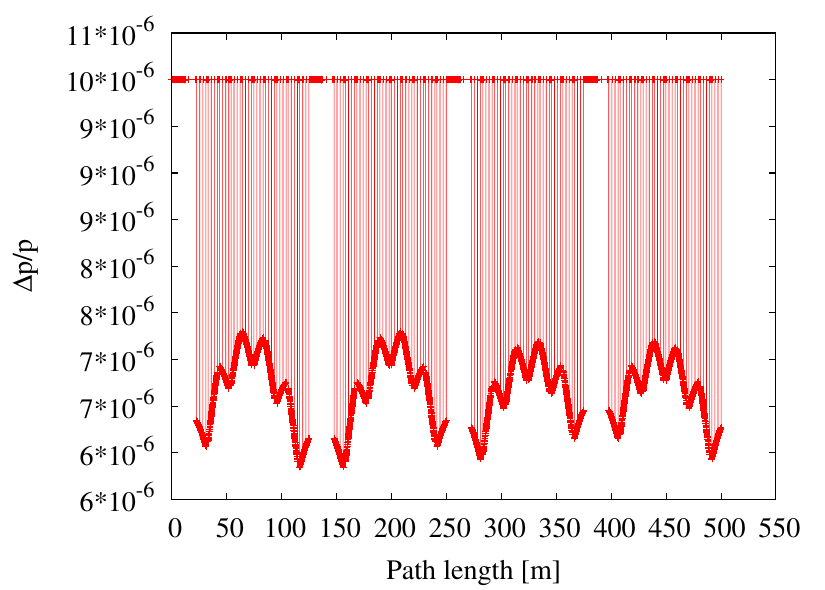}
\caption{Momentum offset as a function of the path length for one turn in the ring. Here $\mean{\Delta p /p} = 7.6*10^{-6}$. Note that the initial momentum offset of the particle, i.e. at injection, is $10^{-5}$.}
\label{moment_spread}
\end{figure}

If the particle is injected with an initial momentum offset, then a vertical spin buildup will occur in presence of vertical motion. Assuming a net vertical misalignment of one quadrupole, one can search for the closed orbit ensuring the periodicity of the ring as illustrated in \mbox{fig \ref{vert_CO}}.
For instance, let's consider the case of a particle injected with a momentum offset $\Delta p/p = 10^{-5}$. Given that the particle is traveling through electrostatic elements, the total energy conservation is a crucial aspect of the simulation. The effect is most dominant inside the electrostatic bends \cite{mane} leading to a non negligible change of the momentum offset of the particle as shown in fig \ref{moment_spread}.
The latter can be calculated in the following way:
\begin{eqnarray}
\dfrac{\Delta p}{p}=\dfrac{\Delta E}{\beta_l c p} = \dfrac{\Delta E_{out}- e\phi}{\beta_l c p} &\approx & \dfrac{1}{\beta_l c p} \left[\Delta E_{out}+ e E_{r}^b \left(x-\dfrac{x^2}{2R_0}\right)- \dfrac{eG}{2}(x^2-y^2)\right]  \nonumber \\
&=& \left(\dfrac{\Delta p}{p}\right)_{out} + \dfrac{eE_r^b}{\beta_l c p}x - \dfrac{eE_r^b/(2R_0)+eG/2}{\beta_l c p}x^2+\dfrac{eG/2}{\beta_l c p}y^2 \nonumber \\
&\approx & \left(\dfrac{\Delta p}{p}\right)_{out} + \dfrac{eE_r^b}{\beta_l c p}x
\label{dp_potential}
\end{eqnarray}
where the subscript $out$ denotes the value outside the electrostatic elements. From what preceded, one can calculate the average spin precession components and one obtains:
\begin{eqnarray}
\mean{\Omega} = \sqrt{\mean{\Omega_y}^2+\mean{\Omega_l}^2} = \sqrt{(-24.09)^2+(9.22*10^{-3})^2} \approx 24.09 \text{ rad/s} 
\end{eqnarray}
so that the spin precession is dominated by the average of its vertical component $\mean{\Omega_y}$. It results that the non-linear terms cannot be neglected, even for timescales of the order of few milliseconds. Thus, relying on the first order approximation (\ref{xi}), one can compute an approximate solution to the vertical spin buildup:
\begin{eqnarray}
s_r(t) &\approx & \dfrac{\mean{\Omega_y}}{\mean{\Omega}} \sin \left(t \mean{\Omega}\right) \approx -\sin \left(24.09*t \right) \label{eq:sr}\\
s_l(t) &\approx & 1+\frac{\mean{\Omega_y}^2}{\mean{\Omega}^2} \left[\cos \left(t \mean{\Omega} \right)-1 \right] \approx \cos \left(24.09*t \right) \label{eq:sl} \\
s_y(t) &\approx & -\frac{\mean{\Omega_l}  \mean{\Omega_y}}{\mean{\Omega}^2} \left[\cos \left(t \mean{\Omega}\right)-1 \right] \approx -7.6348*10^{-4} \left[\cos(24.09*t)-1\right] \label{eq:sy}
\end{eqnarray}

\begin{figure}%
    \centering
    \subfloat[Longitudinal spin versus time.]{{\includegraphics[width=8cm]{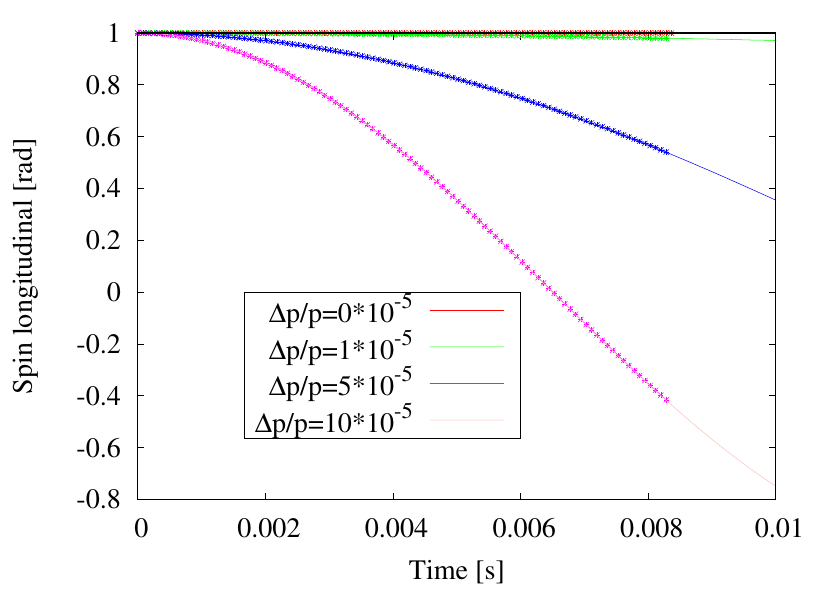} }}%
    \qquad
    \subfloat[Radial spin versus time.]{{\includegraphics[width=8cm]{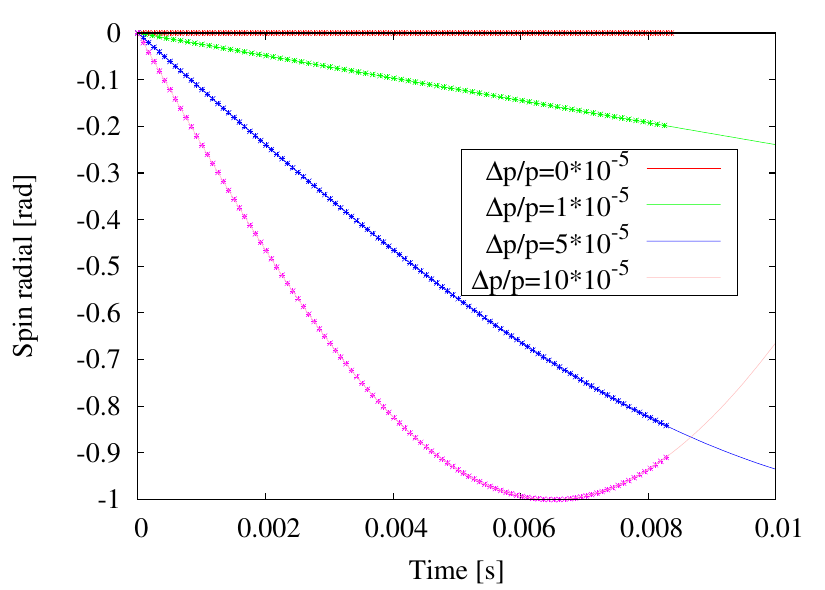} }}%
    \caption{Comparison of the tracking simulations with the analytical estimates (solid lines) for the case with one vertically misaligned quadrupole $\Delta y=10 \text{ $\mu m$}$.}%
    \label{fig:srad_long}%
\end{figure}

Then, several particles are tracked with different initial momenta. Figs \ref{fig:srad_long} shows a comparison of the longitudinal and radial spin results (obtained from BMAD tracking) with the analytical formula (\ref{eq:sr}) and (\ref{eq:sl}). As can be seen, the agreement is good.

\begin{figure}%
    \centering
    \subfloat[Spin buildup for various momentum offsets ($\Delta y =10 \text{ $\mu$m}$).]{{\includegraphics[width=8cm]{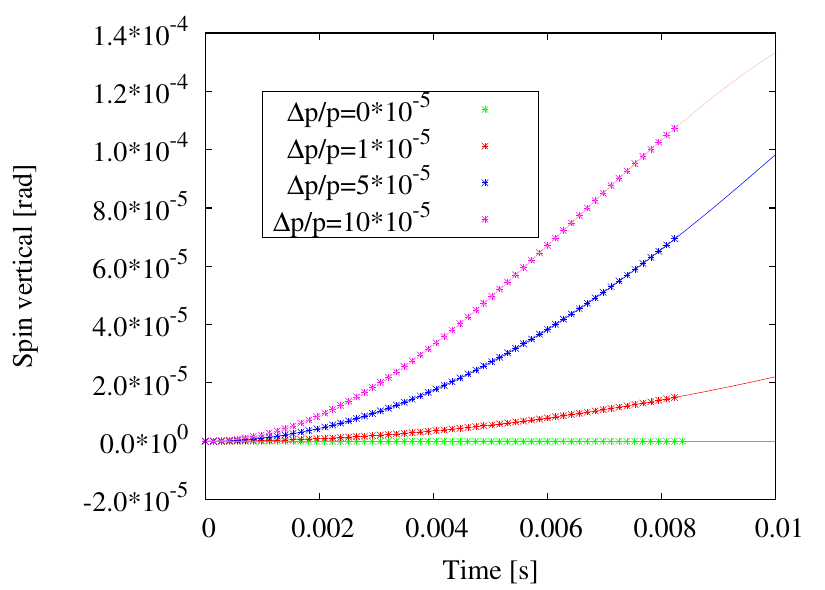} }}%
    \qquad
    \subfloat[Spin buildup for various misalignment erros ($\Delta p/p =10^{-5}$).]{{\includegraphics[width=8cm]{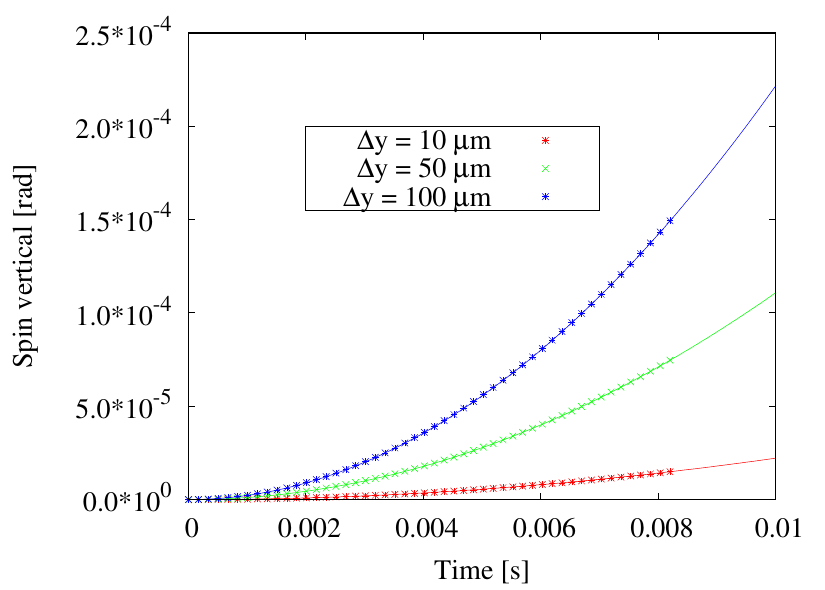} }}%
    \caption{Comparison of the tracking simulations with the analytical estimates (solid lines, Eq. (\ref{eq:sy})) for the case with one vertically misaligned quadrupole.}%
    \label{fig:svert_mis}%
\end{figure}

Next, one varies the initial momentum offset as well as the quadrupole misalignment to determine its impact on the vertical spin buildup. From Eq. (\ref{eq:sy}), a Taylor expansion yields:
\begin{eqnarray}
s_y(t) &=& \dfrac{\mean{\Omega_y} \mean{\Omega_l}}{2} t^2 + O(t^4) \nonumber \\
	   &\propto & \mean{y' E_r} \mean{E_r \Delta p/p}
\end{eqnarray}
so that the vertical spin buildup is proportional to the momentum offset from the magic energy as well as to the vertical slope in the bends and therefore to the net vertical misalignment of the quadrupole. This is confirmed by tracking simulations as shown in fig \ref{fig:svert_mis}. Nevertheless, the first order approximation fails at the magic energy, i.e. when $\Delta p/p \rightarrow 0$ since then, $\mean{\Omega} \rightarrow 0$ and one needs to exploit the higher order terms in order to explain any spin buildup. This is discussed in the next section.

\textbf{NB:} In order to evaluate the above average quantities properly, one has to calculate the averages of all quantities by integrating over the closed orbit. For instance, a particle off-momentum yields a closed orbit that is different from the ideal one. Thus, a search for the closed orbit is necessary in order to avoid any fluctuations of the calculated quantities, since the above formalism is based on the assumption that all quantities are periodic so that the average value is unchanged from one turn to the next. \\

\subsubsection{Case of a beam injected at the magic momentum}
\begin{figure}%
    \centering
    \subfloat[Case where the vertical displacement of both quadrupoles is varied.]{{\includegraphics[width=8cm]{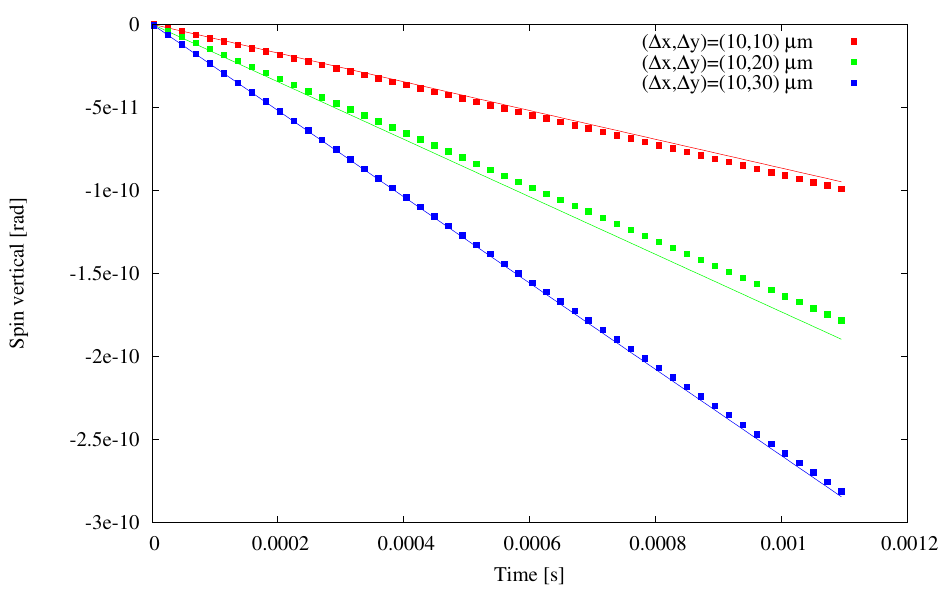} }}%
    \qquad
    \subfloat[Case where the horizontal displacement of both quadrupoles is varied.]{{\includegraphics[width=8cm]{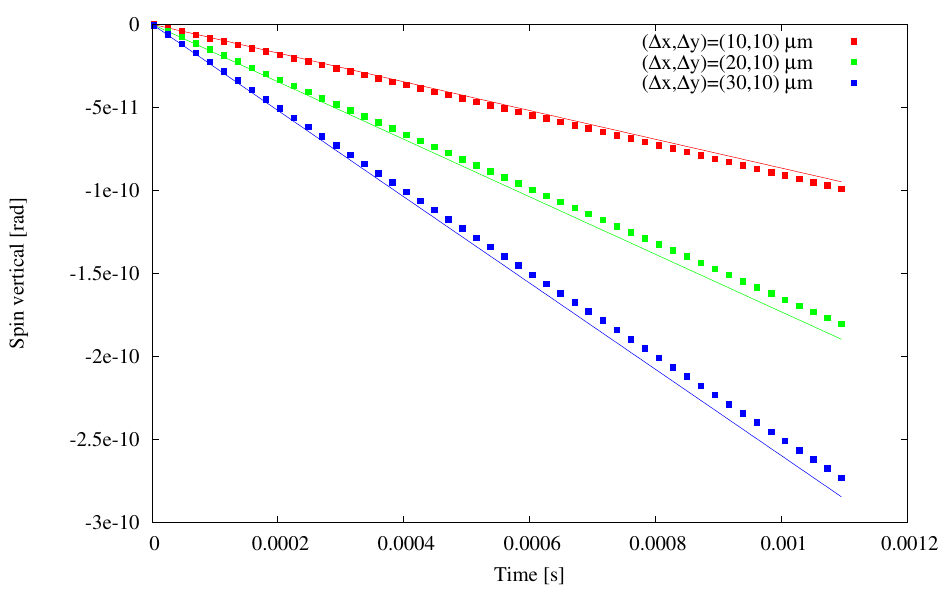} }}%
    \caption{Vertical spin buildup from tracking simulations and comparison with the analytical estimates (solid lines) for the case with alternating misaligned quadrupoles.}%
    \label{fig:defoc_quad}%
\end{figure}

If a beam is injected at the magic momentum, then, in presence of misalignment errors, spin precession will occur. To show this, let's consider a lattice that has two misaligned quadrupoles as follows: in the first quarter of the ring, a defocusing quadrupole is misaligned vertically and horizontally by $(+\Delta x,+\Delta y)$. In the third quarter, i.e. 180 degrees out of phase, a second defocusing quadrupole is misaligned by $(-\Delta x,-\Delta y)$. Thus, the average misalignment vanishes in this configuration. For a beam injected at the magic energy, an energy spread will appear at the entrance of the misaligned quadrupole. This results in a change of the closed orbit and consequently in a change of the kinetic energy within the electrostatic bends:
\begin{eqnarray}
\dfrac{\Delta p}{p} = \dfrac{eE_r^b}{\beta_l c p} x
\end{eqnarray}
This consequently yields a radial spin buildup within the electrostatic bends where the effect is most important, which generates a vertical spin as well by means of a longitudinal spin precession. Assuming \mbox{$\Delta x=\Delta y = 10 \mu m$}, one obtains:
\begin{eqnarray}
s_r(t) &\approx &  \mean{\Omega_y} t + \mean{\Omega_l \tilde{\Omega}_r} t - \mean{\Omega_r} \mean{\tilde{\Omega}_l} t + \dfrac{\mean{\Omega_r} \mean{\Omega_l}}{2} t^2 \nonumber \\
&\approx & -7.11*10^{-4} t -2.66*10^{-7} t + 0*t + 0* t^2
\end{eqnarray}
and
\begin{eqnarray}
s_y(t) &\approx &  -\mean{\Omega_r} t + \mean{\Omega_l \tilde{\Omega}_y} t - \mean{\Omega_y} \mean{\tilde{\Omega}_l} t + \dfrac{\mean{\Omega_y} \mean{\Omega_l}}{2} t^2 \nonumber \\
&\approx & 0 *t -8.68*10^{-8} t + 2.60*10^{-10} t + 1.67*10^{-12} t^2
\end{eqnarray}
Thus, the vertical spin buildup is mainly due to the geometric phases given by:
\begin{eqnarray}
\dfrac{ds_y}{dt} \approx \mean{\Omega_l \tilde{\Omega}_y} &\approx & - \dfrac{1}{C} \int_{L_{bend}} \Omega_l \tilde{\Omega}_y ds \nonumber \\
&\propto &  {E_r}^2 \dfrac{\Delta p}{p} y' \propto \Delta x * \Delta y
\end{eqnarray}
which is proportional to the product of the displacements of both quadrupoles: the horizontal displacement of the quadrupoles yields larger radial spin oscillations due to the variation of the kinetic energy in the electrostatic bends while the vertical displacement of the quadrupoles yields a vertical slope inside the electrostatic bends, therefore a longitudinal spin precession which rotates the radial spin into the vertical plane. Such an effect yields a non-vanishing average value, therefore the frozen spin is proportional to both displacements as verified by tracking simulations in fig \ref{fig:defoc_quad}. 

\clearpage

\section{Magnetic field imperfections and EDM effect}
In the case where the magnetic fields as well as the EDM signal are taken into account, the previous formalism holds, the only difference being the coefficients of the transfer matrix. The spin vector is still propagated with respect to the local reference frame of the accelerator. In other words, the presence of magnetic fields is considered as a perturbation that does not modify the reference frame: the latter remains attached to the ideal orbit in the median plane of the accelerator and although the horizontal restoring force contains a perturbation induced by the transverse magnetic fields, the accelerator reference frame does not rotate with such a perturbation. Such definition shall be taken with care though: in some derivations, one subtracts the angular velocity induced by the vertical magnetic fields $eB_y/(\gamma m)$. Given that the bending is only achieved by the electrostatic deflectors, we do not make such a correction in our calculations. Besides, successive quadrupoles in the lattice provide alternate bending in opposite directions to keep the beam stable so that, on average, such a correction may be neglected. An excellent note discussing this matter can be found in \cite{luccio}. \\
Keeping terms up to the second order only, Eq. (\ref{quant}) re-writes:

\begin{equation}
\begin{cases}
\Omega_y = K - \dfrac{e}{mc} \left[a+\dfrac{1}{\gamma+1} \right] \dfrac{\beta_l}{1 + x/\rho} x' E_l - \dfrac{e}{m} \left(a + \dfrac{1}{\gamma} \right) B_y + \dfrac{e}{m}\dfrac{a}{1+x/\rho} \left(1-\dfrac{1}{\gamma}\right) y'B_l \\ \\
\hspace{8mm} - \dfrac{e}{mc}\dfrac{\eta}{2} \left[E_y - \dfrac{\gamma}{\gamma + 1} (\bm{\beta}.\bm{E}) \beta_l y' +c \beta_l ( B_x - x' B_l) \right] \\ \\
\Omega_l = \dfrac{e}{mc} \left[a+\dfrac{1}{\gamma+1} \right] \dfrac{\beta_l}{1 + x/\rho} (x' E_y-y' E_r) - \dfrac{e}{m} \dfrac{1+a}{\gamma} B_l + \dfrac{e}{m}\dfrac{a}{1+x/\rho} \left(1-\dfrac{1}{\gamma}\right) (x'B_x + y'B_y) \\ \\
\hspace{8mm} - \dfrac{e}{mc} \dfrac{\eta}{2} \left[E_l - \dfrac{\gamma}{\gamma + 1} (\bm{\beta}.\bm{E}) \beta_l + c \beta_l(x' B_y - y' B_x) \right] \\ \\
\Omega_r = \dfrac{e}{mc} \left[a+\dfrac{1}{\gamma+1} \right] \beta_l \left(\dfrac{y'}{1 + x/\rho} E_l-E_y \right) - \dfrac{e}{m} \left(a + \dfrac{1}{\gamma}\right) B_x  + \dfrac{e}{m}\dfrac{a}{1+x/\rho} \left(1-\dfrac{1}{\gamma}\right) x'B_l \\ \\
\hspace{8mm} - \dfrac{e}{mc} \dfrac{\eta}{2} \left[E_r - \dfrac{\gamma}{\gamma + 1} (\bm{\beta}.\bm{E}) \beta_l x' + c \beta_l (y' B_l - B_y) \right] \\ \\
K = \left\{
  \begin{array}{@{}ll@{}}
    \dfrac{-e/mc}{ a \left(\gamma_m +1 \right)^2 \gamma_m} \beta_l E_r^b \dfrac{\Delta p}{p} + \dfrac{e}{mc} \left[a + \dfrac{1}{\gamma + 1} \right] \beta_l (E_r -E_r^b)  - \beta_l c \dfrac{x}{\rho ^2} & \text{if bend} \\ \\
    \dfrac{e}{mc} \left[a+\dfrac{1}{\gamma+1} \right] \beta_l E_r & \text{otherwise}
  \end{array}\right.
\end{cases}
\end{equation}
In what follows, one focuses on investigating the impact of the systematic errors originating from magnetic field imperfections. To simplify the analysis, one assumes a vanishing EDM signal, i.e. $\eta =0$. Unless otherwise specified, one relies on the second order approximation for the benchmarking of the analytical results with the tracking simulations. 

\subsection{Geometric phases contributing to the vertical spin buildup}
If the beam is injected at the magic energy, the geometric phase effects may become important. It is extremely difficult  though to reach such configuration where \mbox{$\mean{\Omega_r}=\mean{\Omega_y}=\mean{\Omega_l}=0$}. In general, one may reach a configuration where only the averages of two components of the spin precession vector vanish. The remaining component can however play an important role to transfer the spin buildup to another plane as is discussed in section \ref{sectionBlBr}.
Table \ref{table:geometric_phases} summarizes the leading terms of the geometric phases as established in \mbox{Eq. (\ref{Eq:geom_phases}).} Some of these contributions, particularly the third order terms cannot be eliminated by using counter-rotating beams. However, if the ring is properly shielded such as the integrated magnetic fields imperfections are below the nT.m level, then one can reasonably neglect such contributions to the vertical spin buildup in comparison with the aimed sensitivity of the experiment ($\sim nrad/s$ EDM signal). Precise orbit control is a condition \textit{sine qua non} to eliminate such contributions mimicking the EDM signal.
\renewcommand{\arraystretch}{2.0}
\begin{table*}
\centering	
\begin{tabular}{|l|c|r|} 
   \hline
    Geometric phases (leading terms) & \multicolumn{2}{c|}{Acting fields} \\
    \hhline{|=|=|=|}
    $\mean{\Omega_l \tilde{\Omega}_y} t$ & $\Omega_l \propto \left(y'E_r, x'E_y, y'B_y, x'B_x, B_l \right)$ & $\Omega_y \propto \left(E_r \Delta p/p, x'E_l, y'B_l, B_y \right)$ \\
    \cline{2-3} 
        &  \multicolumn{2}{c|}{ Leading terms $\propto y'E_r^2 \Delta p/p, y'E_r B_y, B_l B_y$ } \\
    \hhline{|=|=|=|}
    $\mean{\Omega_l \widetilde{\Omega_l \tilde{\Omega}_r}} t - \mean{\Omega_l \tilde{\Omega}_r} \mean{\tilde{\Omega}_l}t$ & $\Omega_l \propto \left(y'E_r, x'E_y, y'B_y, x'B_x, B_l \right)$ & $\Omega_r \propto \left(E_y, y'E_l, x'B_l, B_x \right)$ \\
    \cline{2-3} 
        &  \multicolumn{2}{c|}{ Leading terms $\propto (y'E_r)^2 B_x, B_l^2 B_x$} \\
    \hhline{|=|=|=|}     
    $\dfrac{\mean{\Omega_r (\tilde{\Omega}_y)^2}}{2} t$ & $\Omega_r \propto \left(E_y, y'E_l, x'B_l, B_x \right)$ & $\Omega_y \propto \left(E_r \Delta p/p, x'E_l, y'B_l, B_y \right)$ \\
    \cline{2-3} 
        &  \multicolumn{2}{c|}{Leading terms $\propto (E_r \Delta p/p)^2 B_x, B_y^2 B_x$  } \\
    \hline
\end{tabular}
\caption{Leading terms of the geometric phases and contributing fields.}
\label{table:geometric_phases}
\end{table*}



\subsection{Horizontal magnetic field imperfections}

The particle equation of motion in the vertical plane may be written:
\begin{eqnarray}
\dfrac{dp_y}{dt} &=& e \left(E_y+ v_l B_x - v_x B_l\right) \nonumber \\
&=& e \left(E_y+ \beta_l c (B_x - \dfrac{x'}{1+x/\rho} B_l) \right)
\end{eqnarray}
which yields after integration
\begin{eqnarray}
\dfrac{1}{e} \left[p_y(t) - p_y(0)\right] &=& \int_{0}^{t} E_y(\tau) d\tau + \int_{0}^{t} \beta_l c B_x(\tau) d\tau - \int_{0}^{t} \beta_l c \dfrac{x'}{1+x/\rho} B_l(\tau) d\tau \nonumber \\
&=&\dfrac{\beta_l \gamma mc}{e} \left[y'(t)-y'(0) \right]
\end{eqnarray}
which vanishes on the closed orbit. This implies:
\begin{eqnarray}
\mean{E_y + \beta_l c B_x} = \beta_l c \mean{\dfrac{x'}{1+x/\rho} B_l} \label{meanEB}
\end{eqnarray}
Now, integrating the expression of the radial spin precession, and assuming no longitudinal electric fields one obtains:
\begin{eqnarray}
\int_0^t \Omega_r d\tau &=& -\dfrac{e}{mc} \left(a+\dfrac{1}{\gamma+1} \right) \beta_l \int_0^t E_y d\tau - \dfrac{e}{m} \left(a + \dfrac{1}{\gamma}\right) \int_0^t B_x d\tau + \dfrac{e}{m} a \left(1-\dfrac{1}{\gamma}\right) \int_0^t \dfrac{x'}{1+x/\rho} B_l d\tau \nonumber \\
&=& -\dfrac{e}{mc \beta_l \gamma} \int_0^t \left[ \left(a+\dfrac{1}{\gamma+1} \right) \beta_l^2 \gamma E_y + \left(a + \dfrac{1}{\gamma}\right) \gamma \beta_l c B_x \right] d\tau + \dfrac{e}{m} a \left(1-\dfrac{1}{\gamma}\right) \int_0^t \dfrac{x'}{1+x/\rho} B_l d\tau \nonumber \\
&=& -\dfrac{e}{mc \beta_l \gamma} \int_0^t \left[ E_y + \beta_l c B_x + a \gamma \beta_l c B_x \right] d\tau + \dfrac{e}{m} a \left(1-\dfrac{1}{\gamma}\right) \int_0^t \dfrac{x'}{1+x/\rho}B_l d\tau \label{omegarinteg}
\end{eqnarray}
given that, at the magic energy $\left(a+ \dfrac{1}{\gamma + 1}\right) \beta_l^2 \gamma = 1$. 
Now, injecting Eq. (\ref{meanEB}) into (\ref{omegarinteg}) to compute the average on the closed orbit yields:
\begin{eqnarray}
\int_0^t \Omega_r d\tau &=& -\dfrac{e}{m}a \int_0^t  B_x d\tau - \dfrac{e}{m\gamma} \int_0^t  \dfrac{x'}{1+x/\rho} B_l d\tau + \dfrac{e}{m} a \left(1-\dfrac{1}{\gamma}\right) \int_0^t \dfrac{x'}{1+x/\rho} B_l d\tau \nonumber \\
 &\approx & -\dfrac{e}{m}a \mean{B_x} t -  \dfrac{e}{m(\gamma + 1)} \mean{x' B_l} t \nonumber \\
&\approx & -1.7173*10^{8} \mean{B_x} t  - 0.4261*10^8 \mean{x' B_l} t
\end{eqnarray}
where the magnetic fields are expressed in units of Tesla. For an aimed EDM sensitivity of \mbox{$d_p=10^{-29} \text{e.cm}$} corresponding to $\eta = 1.9*10^{-15}$, the resulting vertical spin buildup is 
\begin{eqnarray}
\dfrac{ds_y}{dt} = \mean{\Omega_r} = -\dfrac{e}{mc}\dfrac{\eta}{2} \mean{E_r} = 1.6 \text{ nrad/s} \hspace{4mm};\hspace{4mm} \mean{E_r}=-5.27 \text{ MV}
\end{eqnarray}
so that an average radial magnetic field of $\approx 10 \text{ aT}$ will generate the same signal as the smallest spin precession to be identified \cite{sredm}. \\
In addition, the presence of any longitudinal magnetic fields can cause a vertical spin buildup and needs to be taken into account in the analysis. Therefore clearing the average radial magnetic fields is not a sufficient condition to eliminate the first order linear terms of the systematic errors arising from the radial spin precession.

\subsection{Longitudinal and vertical magnetic field imperfections}

\subsubsection{Case with $\dfrac{\Delta p}{p}=0$}
\begin{figure}
\centering 
\includegraphics*[width=20cm]{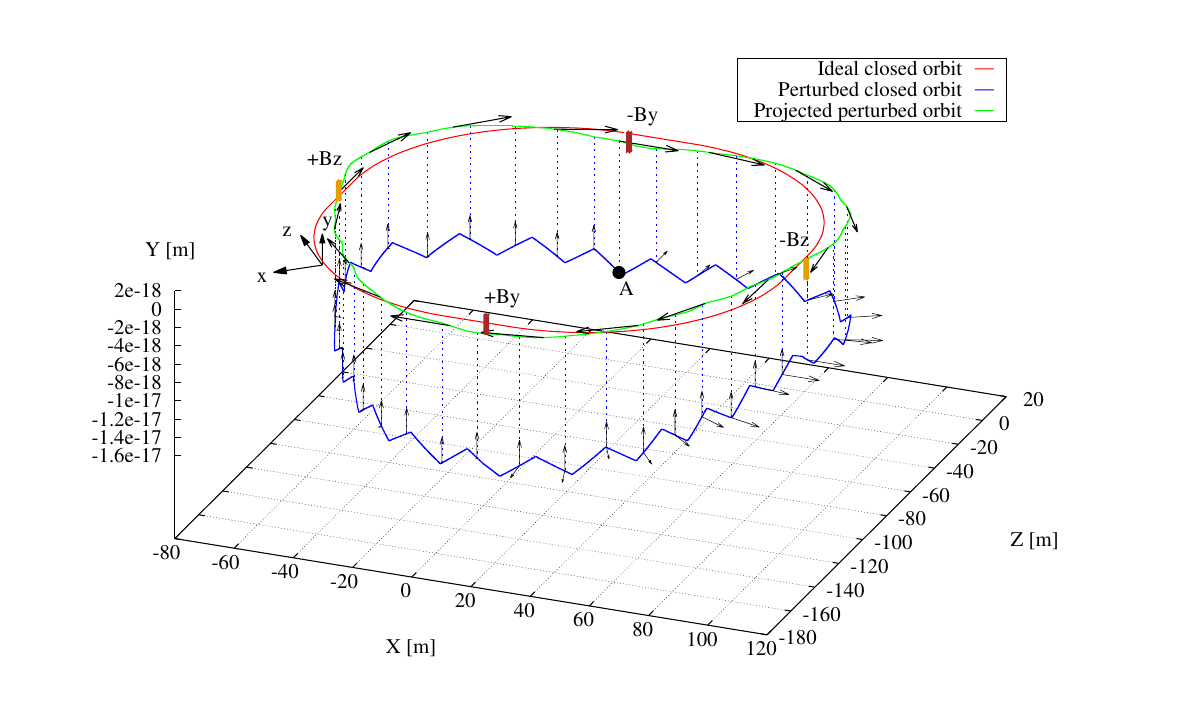}
\caption{Spin and orbit evolution for a lattice with alternating magnetic field imperfections: a vertical magnetic field yields a horizontal spin component which is rotated into the vertical plane by means of a longitudinal field component. The closed orbit of the perturbed motion is shown in blue and the particle motion is clockwise starting from Point A. The orbit displacement from the ideal one is amplified for the sake of clarity.}
\label{orb_3d_sol}
\end{figure}

\begin{figure}
\centering 
\includegraphics*[width=12cm]{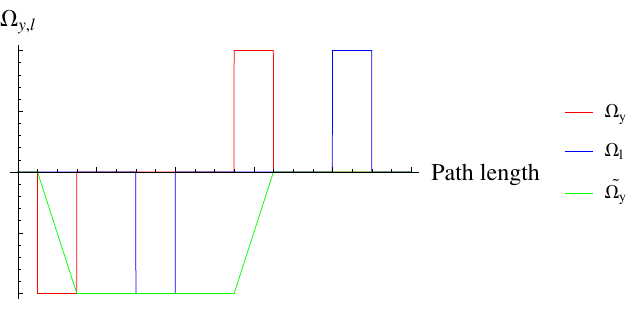}
\caption{Illustrations of the longitudinal and vertical components of the spin precession vector due to alternating longitudinal and vertical magnetic field imperfections. The vertical tilde component $\tilde{\Omega}_y$ represents the integral of the vertical component and accounts for the rapidly oscillating terms of the radial spin component. The average of the product of $\tilde{\Omega}_y$ and $\Omega_l$ yields a non-vanishing vertical spin component.}
\label{omega_illustrate}
\end{figure}

In the configuration where the particle is at the magic energy, the quadratic evolution with time is negligible since the average quantities of the spin precession vectors are negligible. Assuming that the imperfection is due to the presence of both vertical and longitudinal magnetic fields which are $90$ degrees out of phase as illustrated in \mbox{fig \ref{orb_3d_sol}} and for which the integrated fields are \mbox{$\pm 1 \text{ nT.m}$,} one obtains: 
\begin{eqnarray}
s_y(t) &\approx &  -\mean{\Omega_r} t + \mean{\Omega_l \tilde{\Omega}_y} t - \mean{\Omega_y} \mean{\tilde{\Omega}_l} t + \dfrac{\mean{\Omega_y} \mean{\Omega_l}}{2} t^2 \nonumber \\
&\approx & 7.03*10^{-14} t + 2.43*10^{-13} t + 1.22*10^{-15} t -1.47*10^{-20} t^2
\end{eqnarray}
The third term is non vanishing due to an average radial spin precession such as \mbox{$s_r(t) \approx \mean{\Omega_y}t \approx 7.36*10^{-7}t$} which is mainly due to the variation of the energy within the electrostatic deflectors and is proportional to the horizontal orbit offset, therefore to the amplitude of the vertical magnetic fields.
However, the second term is the dominant contribution to the vertical spin buildup. The order of the spin rotations is particularly important to explain the latter as shown below: $\tilde{\Omega}_y$ represents the integral of $\Omega_y-\mean{\Omega_y}$ therefore accounts for the presence of any rapidly oscillating radial spin component. Such alternating magnetic field imperfections in the ring yield alternating spin precession vectors as illustrated in fig \ref{omega_illustrate}. 

Next, one computes the contribution of the Berry phases to the vertical spin buildup:
\begin{eqnarray}
\mean{\Omega_l \tilde{\Omega}_y} = \dfrac{1}{C} \int_{0}^{C} \Omega_l \tilde{\Omega}_y ds
\end{eqnarray}
where $\tilde{\Omega}_y$ is defined by;
\begin{eqnarray}
\tilde{\Omega}_y &=& \int \left[ \Omega_y - \mean{\Omega_y} \right] dt \hspace{2mm};\hspace{2mm} dz= \beta_l c dt \nonumber \\
&=& \dfrac{1}{\beta_l c} \int_0^s \left[ \Omega_y - \mean{\Omega_y} \right] dz
\end{eqnarray}
so that
\begin{eqnarray}
\mean{\Omega_l \tilde{\Omega}_y} &=& \dfrac{1}{C} \int_{0}^{C} \Omega_l \tilde{\Omega}_y ds \nonumber \\
&=& \dfrac{1}{c\beta_l C} \int_{0}^{C} \Omega_l(s) \left(\int_0^s \left[ \Omega_y - \mean{\Omega_y} \right] dz \right) ds \label{lytilde}
\end{eqnarray}
The spin precession is a result of several localized imperfections in the ring. Using the hard edge model, one can express the perturbations as a sum of box distributions in the following way:
\begin{eqnarray}
\Omega(s) = \sum_{i=1}^{N_{err}} \Omega_i \Pi \left(\dfrac{s-s_i}{L_i}-\dfrac{1}{2} \right)
\end{eqnarray} 
where $L_i$ is the length of the perturbation located at $s_i$ and $\Pi(x)$ is the box distribution defined by:
\[\Pi(x) = \begin{cases} 
      1 & |x|\leq 1/2 \\
      0 & \mbox{otherwise} 
   \end{cases}
\]
The integral of the box distribution is the stepwise function defined by:
\[I(x) = \int \Pi(x) dx =  \begin{cases} 
      0 & x\leq -1/2 \\
      1 & x> 1/2 \\
      x+1/2 & \mbox{otherwise} 
   \end{cases}
\]
so that $\tilde{\Omega}$ can be expressed in the following way:
\begin{eqnarray}
\tilde{\Omega}(s) = \sum_{i=1}^{N_{err}} \Omega_i I \left(\dfrac{s-s_i}{L_i}-\dfrac{1}{2} \right)
\end{eqnarray}

\begin{figure}
\centering 
\includegraphics*[width=10cm]{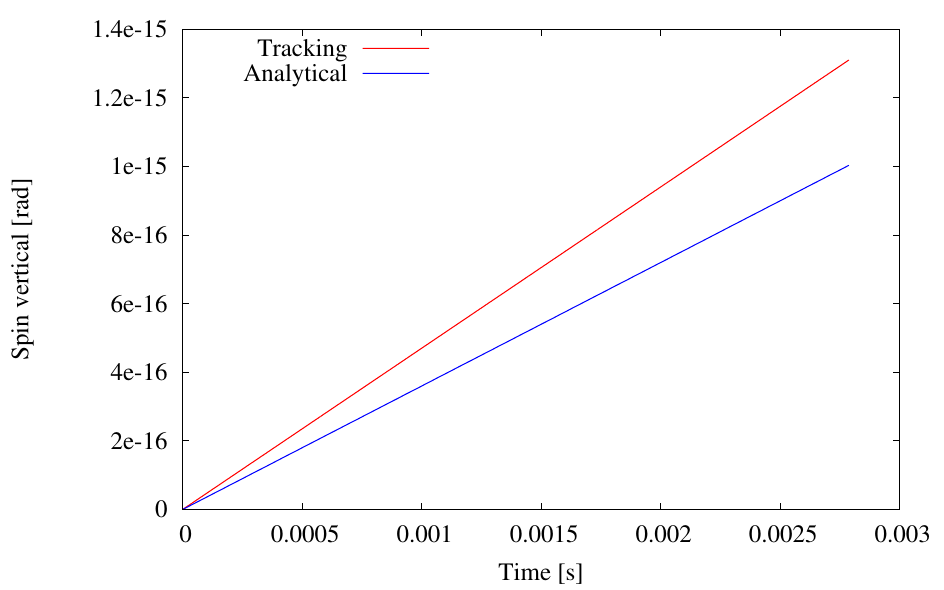}
\caption{Vertical spin buildup from tracking simulations and comparison with the analytical estimate.}
\label{orb_3d_sol2}
\end{figure}

Injecting this into Eq. (\ref{lytilde}) yields:
\begin{eqnarray}
\mean{\Omega_l \tilde{\Omega}_y} &=& \dfrac{1}{c \beta_l C} \sum_{i=1}^{N_l} \sum_{j=1}^{N_y} \Omega_{li} \Omega_{yj} \int_0^C \Pi \left(\dfrac{s-s_i}{L_i}-\dfrac{1}{2} \right)  I\left(\dfrac{s-s_j}{L_j}-\dfrac{1}{2} \right) ds \nonumber \\
&=& \dfrac{1}{c \beta_l C} \sum_{i=1}^{N_l} \sum_{j=1}^{N_y} \Omega_{li} \Omega_{yj} L_i H(s_i-s_j-L_j) 
\end{eqnarray}
where one assumed that both longitudinal and vertical perturbations do not overlap. \mbox{$H$ is} the Heaviside function which accounts for the fact that a vertical spin buildup arises after the radial one.
In the case considered above this yields:
\begin{eqnarray}
\mean{\Omega_l \tilde{\Omega}_y} \approx \dfrac{-L}{c \beta_l C} \left(\dfrac{e}{m}\right)^2 \left(a+\dfrac{1}{\gamma}\right) \dfrac{1+a}{\gamma} B_y B_l L
\end{eqnarray}
which is proportional to the amplitude of the magnetic field perturbations. The tracking simulation results are finally summarized in fig \ref{orb_3d_sol} and \ref{orb_3d_sol2} where one obtained a good agreement.

\subsubsection{Case with $\dfrac{\Delta p}{p} \neq 0$}
\begin{figure}
\centering 
\includegraphics*[width=10cm]{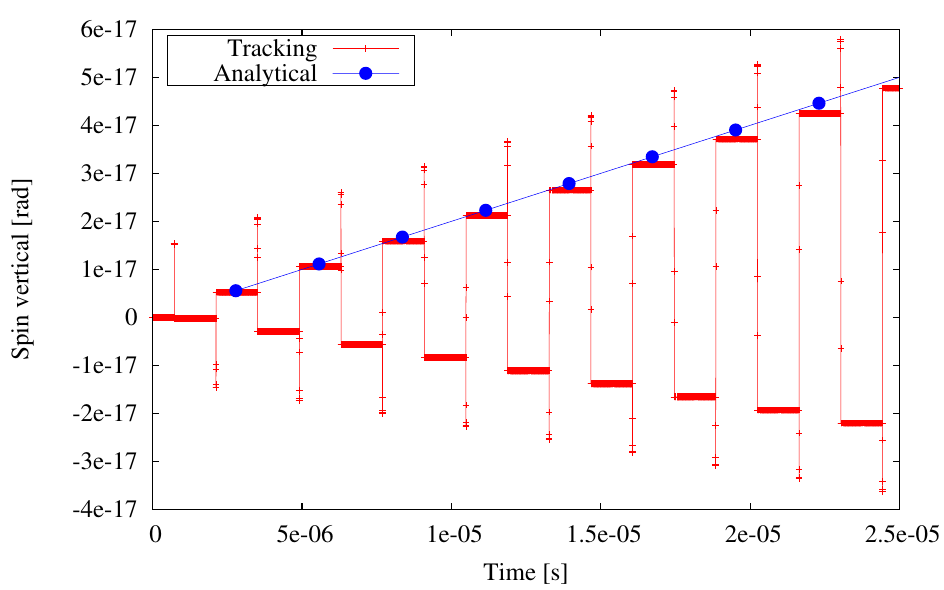}
\caption{Vertical spin buildup from tracking simulations and comparison with the analytical estimate.}
\label{sy_dp}
\end{figure}

In the case where $\Delta p/p = 10^{-9}$, i.e. the particle is slightly above the magic energy, one obtains:
\begin{eqnarray}
s_y(t) &\approx &  -\mean{\Omega_r} t + \mean{\Omega_l \tilde{\Omega}_y} t - \mean{\Omega_y} \mean{\tilde{\Omega}_l} t + \dfrac{\mean{\Omega_y} \mean{\Omega_l}}{2} t^2 \nonumber \\
&\approx & 6.56*10^{-14} t + 2.43*10^{-13} t + 1.66*10^{-12} t -4.96*10^{-18} t^2 \label{analom} \\
&\approx & 1.66*10^{-3} \dfrac{\Delta p}{p} t
\end{eqnarray}
A comparison of the tracking simulations with the analytical estimate (Eq. \ref{analom}) is shown in fig \ref{sy_dp}. The dominant contribution is due to the third term which accounts for the linear slowly varying term of the radial spin component coupled with the rapidly oscillating term of the longitudinal spin precession. \\
In particular, such an effect sets a limit on the difference of energy of the two counter-rotating beams that can be employed to extract the EDM signal: for the case considered here with alternating magnetic field imperfections of 1nT.m, one obtains:
\begin{eqnarray}
\left| \left( \dfrac{\Delta p}{p} \right)_{cw} - \left( \dfrac{\Delta p}{p} \right)_{ccw} \right| < \dfrac{\eta \mean{E_r}}{1.66*10^{-3}} \approx 10^{-6}
\end{eqnarray}

\begin{figure}
\centering 
\includegraphics*[width=10cm]{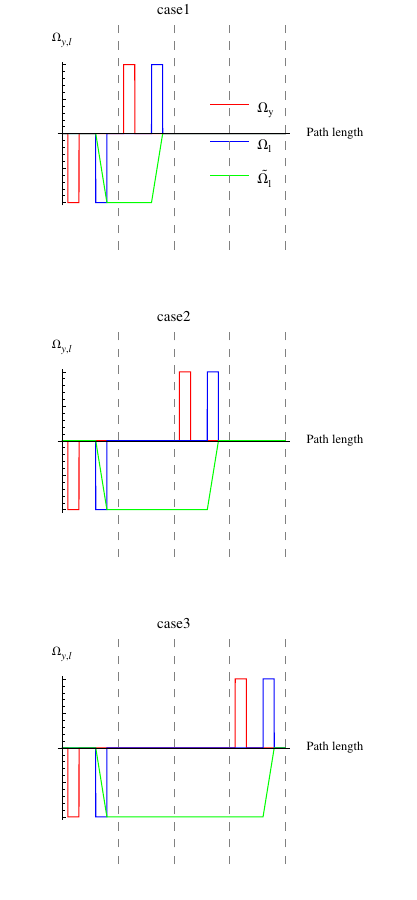}
\caption{Different cases where the alternating field imperfections are located differently in the lattice.}
\label{sy_dp_omegaltilde}
\end{figure}

\begin{figure}
\centering 
\includegraphics*[width=10cm]{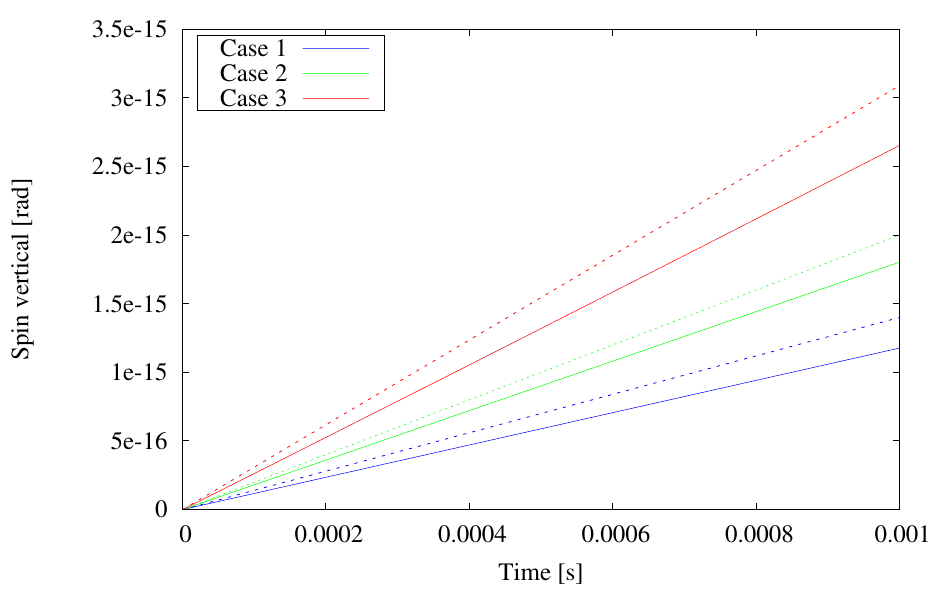}
\caption{Vertical spin buildup from tracking simulations (solid lines) and comparison with the analytical estimate (dashed lines).}
\label{sy_omegaltilde}
\end{figure}

In reality though, the amplitude of the vertical spin buildup is dependent on the location of the perturbation in the lattice, in particular on the shape of the integral of the longitudinal spin precession function, i.e. $\tilde{\Omega}_l$. To show this, let's calculate the latter based on the Heaviside function model:
\begin{eqnarray}
\tilde{\Omega}_l &=& \int \left[ \Omega_l - \mean{\Omega_l} \right] dt \hspace{2mm};\hspace{2mm} \mean{\Omega_l} \approx 0 \nonumber \\
&=& \dfrac{1}{\beta_l c} \int_0^s \left[\sum_{i=1}^n \Omega_{li} \Pi \left(\dfrac{s-s_i}{L_i} - \dfrac{1}{2} \right)  \right] dz \nonumber \\
&=& \dfrac{1}{\beta_l c} \sum_{i=1}^n \Omega_{li} I \left(\dfrac{s-s_i}{L_i} - \dfrac{1}{2} \right)
\end{eqnarray}
Now, assuming that $\Omega_l$ consists of two alternating perturbations at locations $s_1$ and $s_2$ with the same length $L$, one can compute the integral:
\begin{eqnarray}
\mean{\tilde{\Omega}_l} &=& \dfrac{1}{C \beta_l c} \sum_{i=1}^n \Omega_{li} \int_0^C I \left(\dfrac{s-s_i}{L_i} - \dfrac{1}{2} \right) ds \nonumber \\
&=& \dfrac{\Omega_{l1} (s_2-s_1) L}{C \beta_l c}
\end{eqnarray}
which shows that the amplitude of the vertical spin buildup due to the $\mean{\Omega_y}\mean{\tilde{\Omega}_l}$ is proportional to the distance between the two alternating perturbations of the longitudinal magnetic fields.
To verify this result, one constructs several cases where the location of the perturbation is varied as shown in fig \ref{sy_dp_omegaltilde}. Tracking simulations are performed for each case and one can see in fig \ref{sy_omegaltilde} good agreement between the numerical simulations and the analytical estimates.

\subsection{Longitudinal and radial magnetic field imperfections} \label{sectionBlBr}
\begin{figure}
\centering 
\includegraphics*[width=10cm]{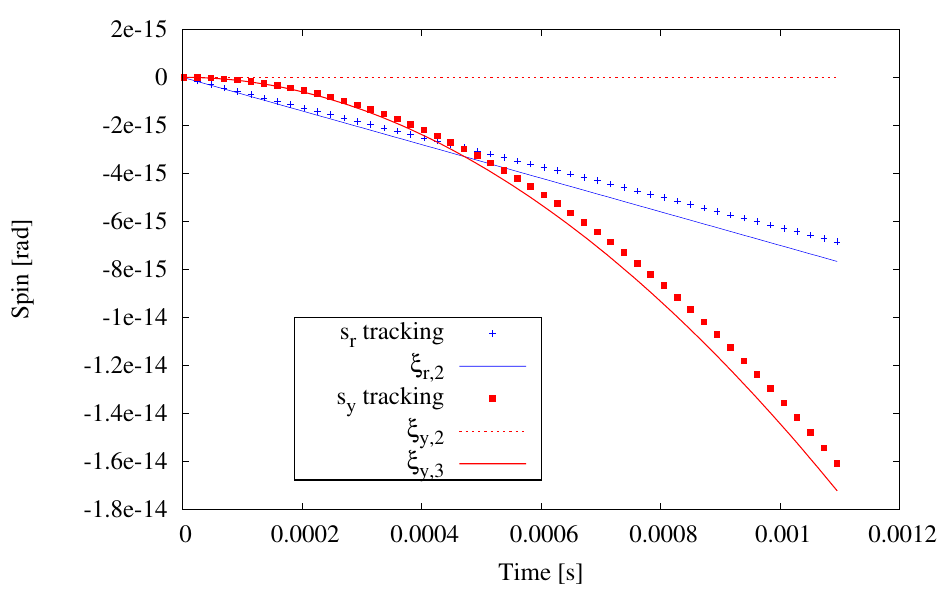}
\caption{Spin buildup and comparison with the analytical estimates. The radial spin is normalized by a factor of 1000 to fit in the same scale. }
\label{SpinBLongRad}
\end{figure}

Assuming that the only perturbation for a particle at the magic energy is due to the radial and longitudinal magnetic fields such that $\mean{\Omega_y},\mean{\Omega_r} \approx 0$ and $\mean{\Omega_l}= 4.3$ rad/s, it results from the second order approximation that the radial and vertical spin buildup are mainly due to the geometric phases such as:
\begin{eqnarray}
\begin{cases}
\xi_{r,2}(t) = \mean{\Omega_y}t + \mean{\Omega_l \tilde{\Omega}_r} t = 1.56 *10^{-9} -8.62 * 10^{-9} \text{rad} \\
\xi_{y,2}(t) = \mean{\Omega_l \tilde{\Omega}_y} t = 1.7 * 10^{-15} \text{rad}
\end{cases}
\end{eqnarray}
\begin{figure}
\centering 
\includegraphics*[width=10cm]{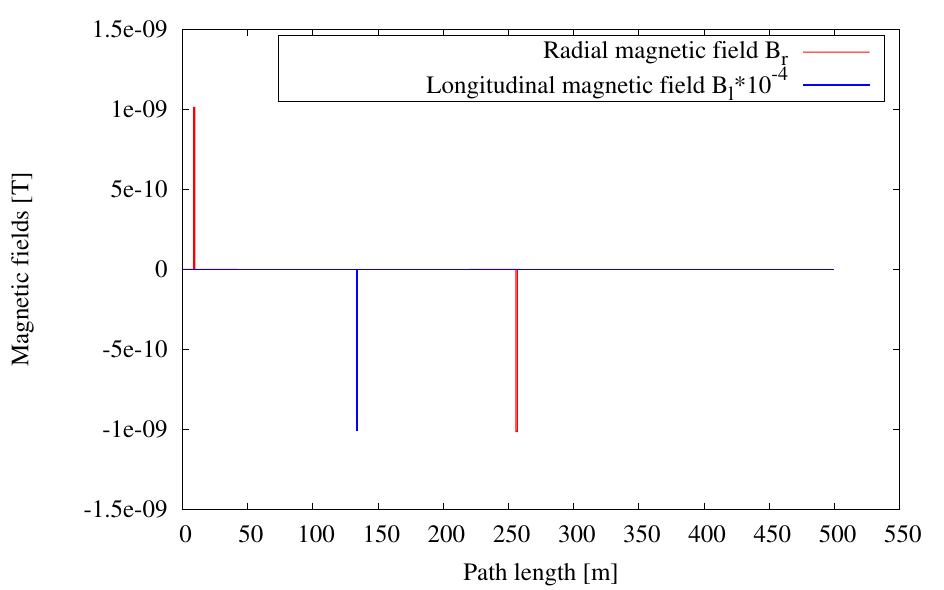}
\caption{Radial and longitudinal magnetic field imperfections in the ring. The integrated fields, in order, correspond to $+1$ nT.m, $-10^4$ nT.m and $-1$ nT.m }
\label{BLongRad}
\end{figure}
The simulated imperfections are shown in fig \ref{BLongRad} and the comparison with the analytical results are summarized in fig \ref{SpinBLongRad}. As can be seen the second order approximation is sufficient to explain the radial spin buildup. However, it fails to explain the quadratic spin buildup in the vertical plane. Thus, the third order approximation is invoked which yields:
\begin{align}
\begin{split}
\xi_{y,3}(t) \approx{} &\left[ \mean{\Omega_l \tilde{\Omega}_y} + \mean{\Omega_l \widetilde{\Omega_l \tilde{\Omega}_r}} - \mean{\Omega_l \tilde{\Omega}_r} \mean{\tilde{\Omega}_l}\right] t \nonumber \\
     &+ \dfrac{\mean{\Omega_l}}{2} \left[\mean{\Omega_y} + \mean{\Omega_l \tilde{\Omega}_r} \right] t^2 + \left[\mean{\Omega_r} \dfrac{\mean{\Omega_r}^2+\mean{\Omega_y}^2}{6}+\dfrac{\mean{\Omega_r}\mean{\Omega_l}^2}{6} \right] t^3 
\end{split} \\
\begin{split}
={}& 2.41*10^{-14}t - 1.58*10^{-8}t^2 + 2.72*10^{-7}t^3
\end{split}
\end{align}

In summary, the geometric phases explain in part the radial spin buildup. However, the latter are transferred to the vertical plane by means of an average longitudinal spin precession.

\subsection{Longitudinal magnetic field imperfections}
\begin{figure}
\centering 
\includegraphics*[width=20cm]{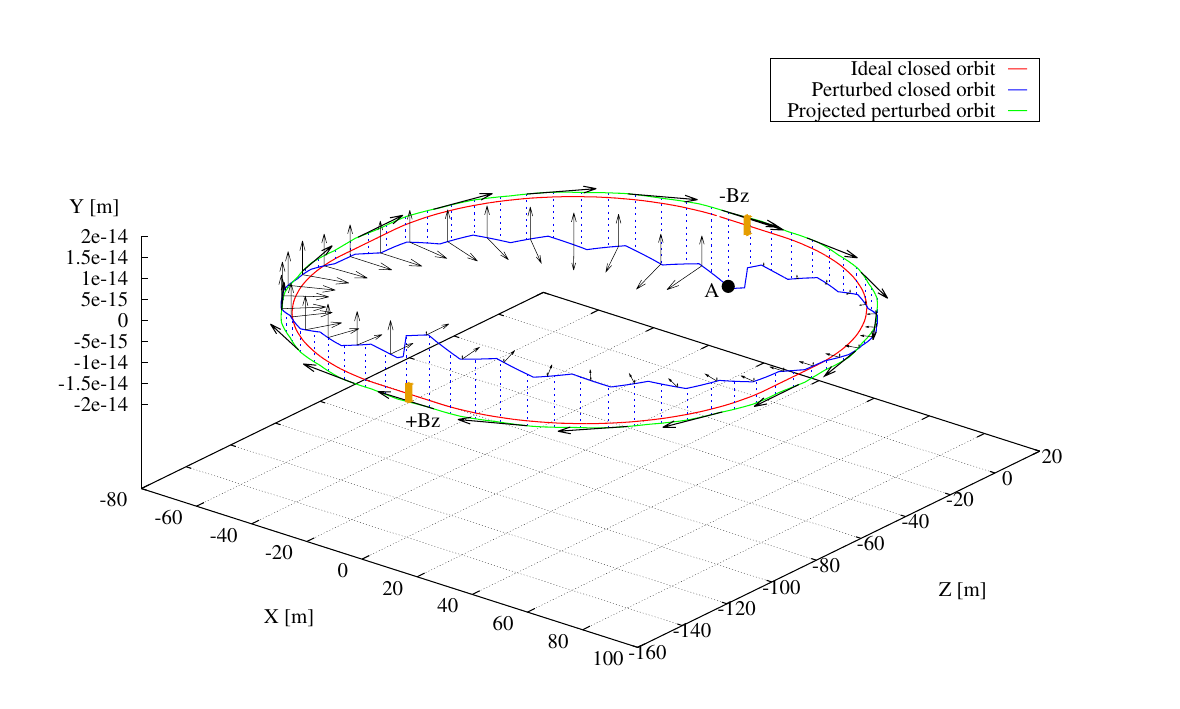}
\caption{Spin and orbit evolution for a lattice with alternating magnetic field imperfections: due to the momentum offset from the magic energy, a radial spin component grows linearly with time. The latter is rotated into the vertical plane by means of a longitudinal magnetic field component. The closed orbit of the perturbed motion is shown in blue and the particle motion is clockwise starting from Point A.}
\label{orb_3d_sol3}
\end{figure}

\begin{figure}%
    \centering
    \subfloat[Alternating longitudinal magnetic field imperfections for two simulated cases with $\Delta p/p=10^{-8}$.]{{\includegraphics[width=8cm]{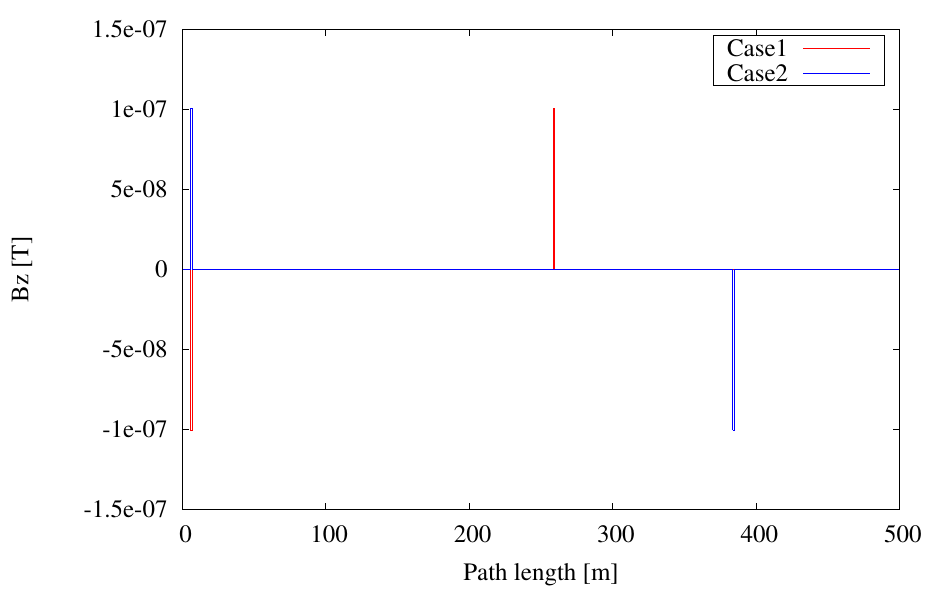} }}%
    \qquad
    \subfloat[The analytical estimates are in solid lines while the tracking simulations are shown with markers. ]{{\includegraphics[width=8cm]{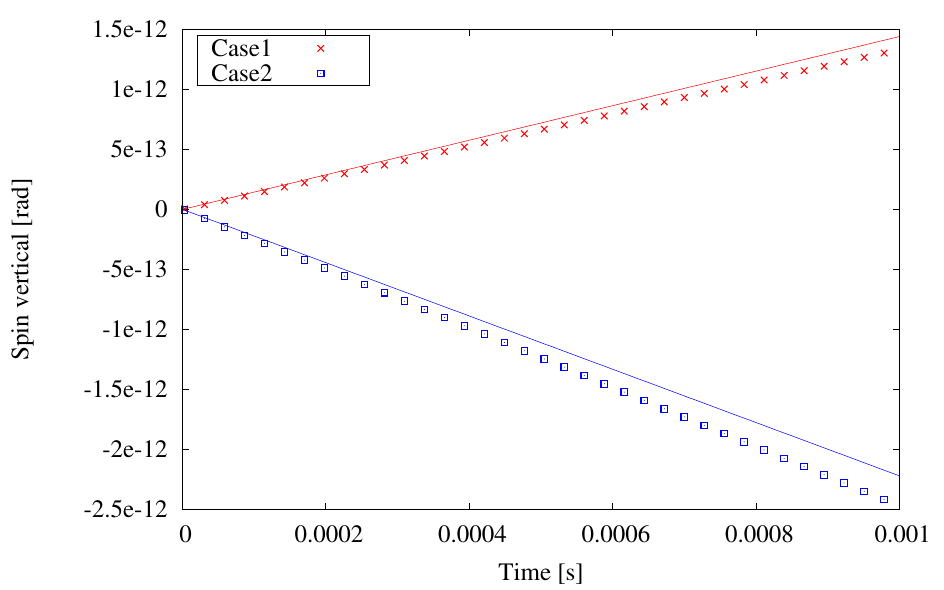} }}%
    \caption{Vertical spin buildup from tracking simulations and comparison with the analytical estimates.}%
    \label{fig:example}%
\end{figure}

Assuming that the only perturbation is due to longitudinal magnetic field imperfections such that $\mean{B_l}=0$, a vertical spin buildup can arise only if the particle is not at the magic energy which yields two contributions: a non-vanishing radial spin precession due to the horizontal orbit displacements coupled with the longitudinal magnetic fields, therefore $ ds_y/dt = \Omega_r s_l \propto \mean{x' B_l} s_l $, and a non vanishing longitudinal spin precession coupled with the radial spin component. For instance, assuming an initial momentum offset $\Delta p/p=10^{-8}$ and $B_l L = \pm 10^{-7} \text{ T.m}$, it follows from \mbox{Eq. (\ref{precise_sysimpl})} that:
\begin{eqnarray}
s_y(t) &\approx &  -\mean{\Omega_r} t + \mean{\Omega_l \tilde{\Omega}_y} t - \mean{\Omega_y} \mean{\tilde{\Omega}_l} t + \dfrac{\mean{\Omega_y} \mean{\Omega_l}}{2} t^2 \nonumber \\
&\approx & 2.35*10^{-12} t - 3.63*10^{-11} t + 1.48*10^{-9} t + 5.86*10^{-17} t^2
\end{eqnarray}
where $\tilde{\Omega}_l$ is given by:
\begin{eqnarray}
\mean{\tilde{\Omega}_l} &=& \dfrac{\Omega_{l1} (s_2-s_1) L}{C \beta_l c} \nonumber \\
&=& \dfrac{-e}{m c \beta_l \gamma} \dfrac{1+a}{C} B_{l1} L (s_2-s_1) \\
&=& -2.38*10^{-3} B_{l1} L (s_2-s_1) 
\end{eqnarray}
Such an effect is illustrated in fig \ref{orb_3d_sol3} where one can see that, starting from point A, a linear radial spin component grows linearly with time. The latter yields a vertical spin buildup which depends on the location of the perturbations in the ring. This is verified by simulating two different lattices with different alternating field imperfections as illustrated in fig \ref{fig:example}.

\begin{figure}
\centering 
\includegraphics*[width=10cm]{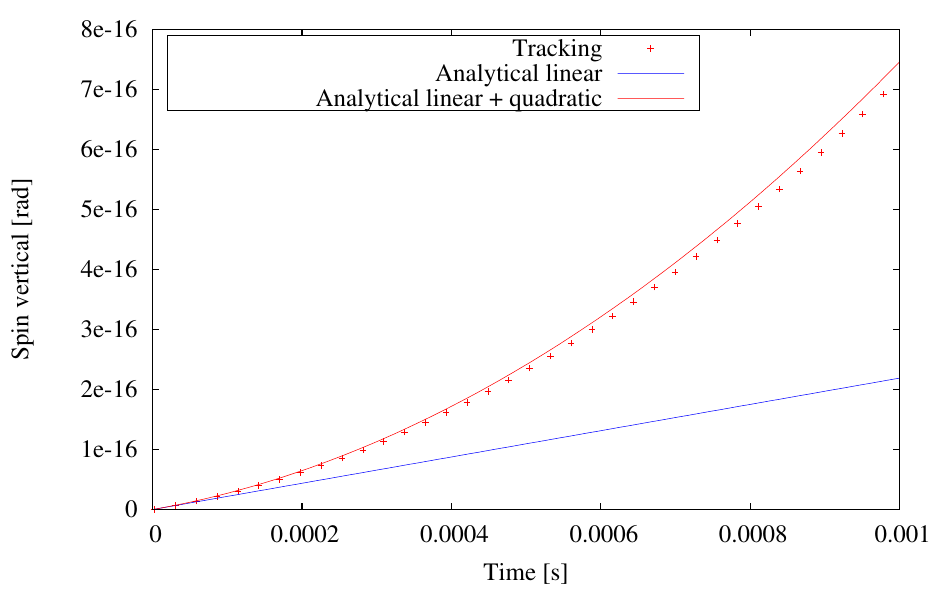}
\caption{Different cases where the alternating field imperfections are located differently in the lattice.}
\label{sy_bz2}
\end{figure}

If the average contribution of the longitudinal magnetic fields is non-vanishing, then a quadratic increase of the vertical spin buildup with time is expected. Assuming $\Delta p/p=10^{-12}$ and an average longitudinal magnetic field of \mbox{$1$ nT.m,} one evaluates the vertical spin buildup by computing the averages:
\begin{eqnarray}
s_y(t) &\approx &  -\mean{\Omega_r} t + \mean{\Omega_l \tilde{\Omega}_y} t - \mean{\Omega_y} \mean{\tilde{\Omega}_l} t + \dfrac{\mean{\Omega_y} \mean{\Omega_l}}{2} t^2 \nonumber \\
&\approx & 1.11*10^{-17} t - 2.93*10^{-15} t + 2.22*10^{-13} t + 5.27*10^{-10} t^2
\end{eqnarray}
Comparison with the tracking simulations is shown in fig \ref{sy_bz2} where one can observe a good agreement between the two calculations.

\section{Conclusion}
Several formula were established and benchmarked against numerical simulations for the purpose of computing the systematic errors in an all-electric proton EDM storage ring. In particular, it appears that the established formalism is very useful to calculate the spin precession rates at the location of the polarimeter, i.e. at specific locations of the storage ring. To reach the sensitivity level required for such an experiment, the second order approximation based on the BKM method of averages is generally sufficient. In addition, it is very practical to benchmark different simulation codes for precision tracking by simply comparing the average quantities of the spin precession components and their periodic tilde functions.

\section*{Acknowledgements}
We acknowledge useful discussions with members of the CPEDM and JEDI collaboration, specifically Mike Lamont, Sig Martin, Selcuk Hac\ifmmode \imath \else \i \fi{}\"omero\ifmmode \breve{g}\else \u{g}\fi{}lu, Andreas Lehrach, Yannis Semertzidis, Ed Stephenson, Hans Stroeher and Richard Talman. Special thanks to David Sagan and Yann Dutheil for helping with BMAD and useful discussions as well.  

\clearpage

\appendix
\section{Spin Tracking example using Mathematica} \label{app:A}
\newcommand{\mathsym}[1]{{}}
\newcommand{\unicode}[1]{{}}

\newcounter{mathematicapage}

\noindent\(\pmb{\text{ClearAll}[\text{{``}Global$\grave{ }$*{''}}]}\) \\ \\
\noindent\(\pmb{\text{SetOptions}[\text{$\$$FrontEndSession},\text{PrintPrecision}\to \text{Ceiling}[\text{$\$$MachinePrecision}]]}\) \\ \\
\noindent\(\pmb{T=2.7872*10^{-6};}\) \\ \\
\noindent\(\pmb{\text{Norder}=7;}\) \\ \\
\noindent\(\pmb{\Omega _l[\text{t$\_$}]=-20*(\text{Sin}[8*\text{Pi}/T*t])-10^{-2};}\) \\ \\
\noindent\(\pmb{\Omega _{\text{la}}=\text{Integrate}\left[\left.\Omega _l[t]\right/T,\{t,0,T\}\right]}\) \\ \\
\noindent\(\pmb{\Omega _{\text{lt}}[\text{x$\_$}]=\text{Integrate}\left[\Omega _l[t]-\Omega _{\text{la}},\{t,0,x\}, \text{Assumptions}\to x\text{$>$=}0
\&\& x<0.01\right];}\) \\ \\
\noindent\(\pmb{\Omega _{\text{lta}}=\text{Integrate}\left[\left.\Omega _{\text{lt}}[t]\right/T,\{t,0,T\}\right]}\) \\ \\
\noindent\(\pmb{\Omega _{\text{ltt}}[\text{x$\_$}]=\text{Integrate}\left[\Omega _{\text{lt}}[t]-\Omega _{\text{lta}},\{t,0,x\}, \text{Assumptions}\to x\text{$>$=}0 \&\& x<0.01\right];}\) \\ \\
\noindent\(\pmb{\Omega _r[\text{t$\_$}]=-2*10^{-10}*\text{Sin}[6*\text{Pi}/T*t];}\) \\ \\
\noindent\(\pmb{\Omega _{\text{ra}}=\text{Integrate}\left[\left.\Omega _r[t]\right/T,\{t,0,T\}\right]}\) \\ \\
\noindent\(\pmb{\Omega _{\text{rt}}[\text{x$\_$}]=\text{Integrate}\left[\Omega _r[t]-\Omega _{\text{ra}},\{t,0,x\}, \text{Assumptions}\to x\text{$>$=}0 \&\& x<0.01\right];}\) \\ \\
\noindent\(\pmb{\Omega _y[\text{t$\_$}]=20*\text{Sin}[10*\text{Pi}/T*t]+\text{Cos}[8*\text{Pi}/T*t]-10;}\) \\ \\
\noindent\(\pmb{\Omega _{\text{ya}}=\text{Integrate}\left[\left.\Omega _y[t]\right/T,\{t,0,T\}\right]}\) \\ \\
\noindent\(\pmb{\Omega _{\text{yt}}[\text{x$\_$}]=\text{Integrate}\left[\Omega _y[t]-\Omega _{\text{ya}},\{t,0,x\}, \text{Assumptions}\to x\text{$>$=}0
\&\& x<0.01\right];}\) \\ \\
\noindent\(\pmb{\Omega _{\text{lyta}}=\text{Integrate}\left[\Omega _l[t]*\left.\Omega _{\text{yt}}[t]\right/T,\{t,0,T\}\right]}\) \\ \\
\noindent\(\pmb{\Omega _{\text{lytt}}[\text{x$\_$}]=\text{Integrate}\left[\Omega _l[t]\Omega _{\text{yt}}[t]-\Omega _{\text{lyta}},\{t,0,x\}, \text{Assumptions}\to
x\text{$>$=}0 \&\& x<0.01\right];}\) \\ \\
*************  second order frozen solution for the radial spin ************** \\ 
*************   \hspace{6mm} $\xi_{r,2}(t) = (r1+r2+r3)*t + r4*t^2$
\\
\noindent\(\pmb{\text{r1}=\Omega _{\text{ya}}}\) \\ \\
\noindent\(\pmb{\text{r2}=\text{Integrate}\left[\Omega _l[t]*\left.\Omega _{\text{rt}}[t]\right/T,\{t,0,T\}\right]}\) \\ \\
\noindent\(\pmb{\text{r3}=-\Omega _{\text{ra}}*\Omega _{\text{lta}}}\) \\ \\
\noindent\(\pmb{\text{r4}=\Omega _{\text{ra}}*\left.\Omega _{\text{la}}\right/2}\) \\ \\
*************  second order frozen solution for the vertical spin ************** \\
*************   \hspace{6mm} $\xi_{y,2}(t) = (t1+t2+t3)*t + t4*t^2$
 \\
\noindent\(\pmb{\text{t1}=-\Omega _{\text{ra}}}\) \\ \\
\noindent\(\pmb{\text{t2}=\Omega _{\text{lyta}}}\) \\ \\
\noindent\(\pmb{\text{t3}=-\Omega _{\text{ya}}*\text{Integrate}\left[\left.\Omega _{\text{lt}}[t]\right/T,\{t,0,T\}\right]}\) \\ \\
\noindent\(\pmb{\text{t4}=\Omega _{\text{ya}}*\left.\Omega _{\text{la}}\right/2}\) \\ \\
\noindent\(\pmb{\Omega _{\text{tot}}=\left(\Omega _{\text{ra}}{}^2+\Omega _{\text{ya}}{}^2+\Omega _{\text{la}}{}^2\right){}^{0.5}}\) \\ \\
\noindent\(\pmb{\text{sol}=\text{NDSolve}[\{s_r'[t]==\Omega _y[t]*s_l[t]-\Omega _l[t]*s_y[t],s_y'[t]==\Omega _l[t]*s_r[t]-\Omega _r[t]*s_l[t],}\) \\
\noindent\(\pmb{s_l'[t]==\Omega _r[t]*s_y[t]-\Omega_y[t]*s_r[t],s_y[0]==s_r[0]==0,s_l[0]==1\}},\) \\
\noindent\(\pmb{\{s_r[t],s_y[t],s_l[t]\},\{t,0,400.1*T\}, \text{MaxStepSize}\to 10^{-9},\text{MaxSteps}\to 8000000,}\) \\
\noindent\(\pmb{\text{Method}\to \{\text{{``}ExplicitRungeKutta{''}},\text{{``}DifferenceOrder{''}}\to
\text{Norder}\}];}\) \\ \\
*************  solution for the vertical spin using explicit Runge Kutta tracker ************** \\ \\
\noindent\(\pmb{\text{sytrack}[\text{t$\_$}]=s_y[t]\text{/.}\text{sol};}\) \\ \\
\noindent\(\pmb{\text{frozsytrack}[\text{t$\_$}]=\text{sytrack}[\text{IntegerPart}[t/T]*T];}\) \\ \\ 
*************  second order analytical solution for the vertical spin ************** \\
*************   \hspace{6mm} $\xi_{y,2}(t) + \phi_{y,2}(t)$ \\ \\
\noindent\(\pmb{\text{syanal}[\text{t$\_$}]=(\text{t1}+\text{t2}+\text{t3})*t+\text{t4}*t^2-\Omega _{\text{rt}}[t]+\Omega _{\text{ya}}*\left(\Omega
_{\text{lt}}[t]*t-\Omega _{\text{ltt}}[t]\right)+\Omega _{\text{lytt}}[t];}\) \\ \\
\noindent\(\pmb{\text{frozsyanal}[\text{t$\_$}]=(\text{t1}+\text{t2}+\text{t3})*t+\text{t4}*t^2;}\) \\ \\
\noindent\(\pmb{\text{plotsy}=\text{Plot}[\{\text{sytrack}[t], \text{syanal}[t],\text{frozsytrack}[t],\text{frozsyanal}[t]\},\{t,0,1.1*T\}]}\) \\ \\
*************  compute and plot the error of the analytical approximation ************** \\ \\
\noindent\(\pmb{\text{error}[\text{t$\_$}]=\text{sytrack}[t]-\text{syanal}[t];}\) \\ \\
\noindent\(\pmb{\text{frozerror}[\text{t$\_$}]=\text{frozsytrack}[t]-\text{frozsyanal}[\text{IntegerPart}[t/T]*T];}\) \\ \\
\noindent\(\pmb{\text{errorplot}=\text{LogLogPlot}[\{\text{Abs}[\text{error}[t]],\text{Abs}[\text{frozerror}[t]]\},\{t,1.1*T,400.1*T\}]}\)

\clearpage

\section{Stiffness studies using explicit Runge Kutta tracker} \label{app:B}
\begin{figure}
\centering 
\includegraphics*[width=10cm]{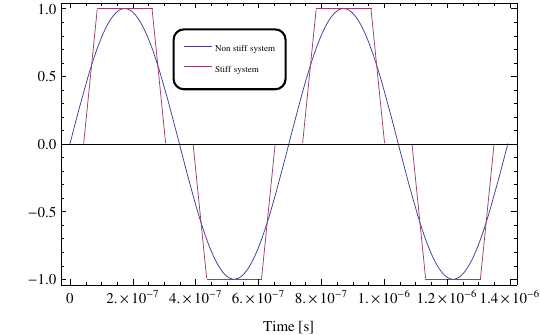}
\caption{Stepwise function approximating the hard edge model. By controlling the width of the rising function, one can control the stiffness of the system. A hard edge model corresponds to a zero width of the rising function.}
\label{stiff}
\end{figure}

\begin{figure*}%
    \centering
    \subfloat[The reference solution corresponds to the $9th$ order Runge Kutta method with the smallest step size.]{{\includegraphics[width=8cm]{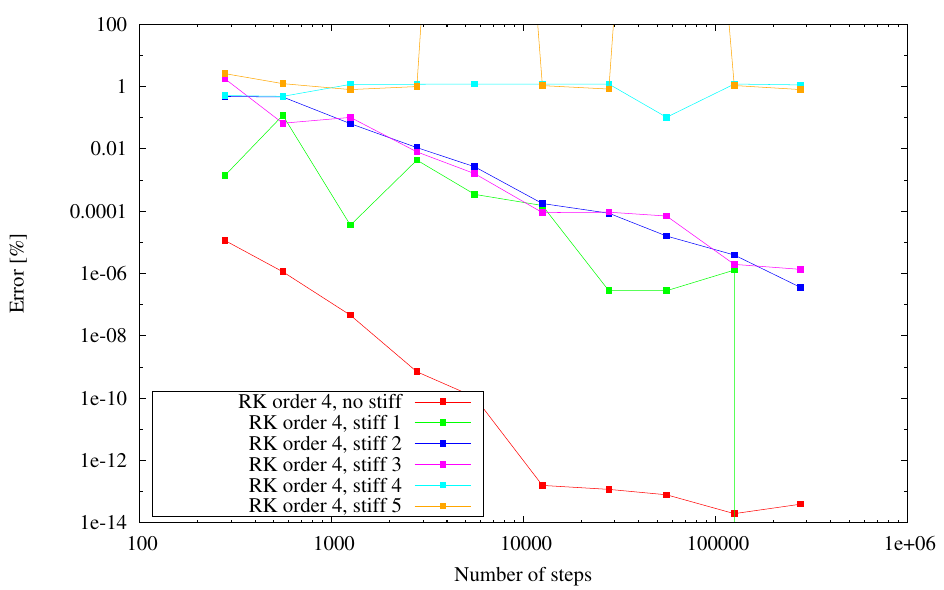} }}%
    \qquad
    \subfloat[The reference solution is the analytical one. Eq. (\ref{precise_sysimpl}).]{{\includegraphics[width=8cm]{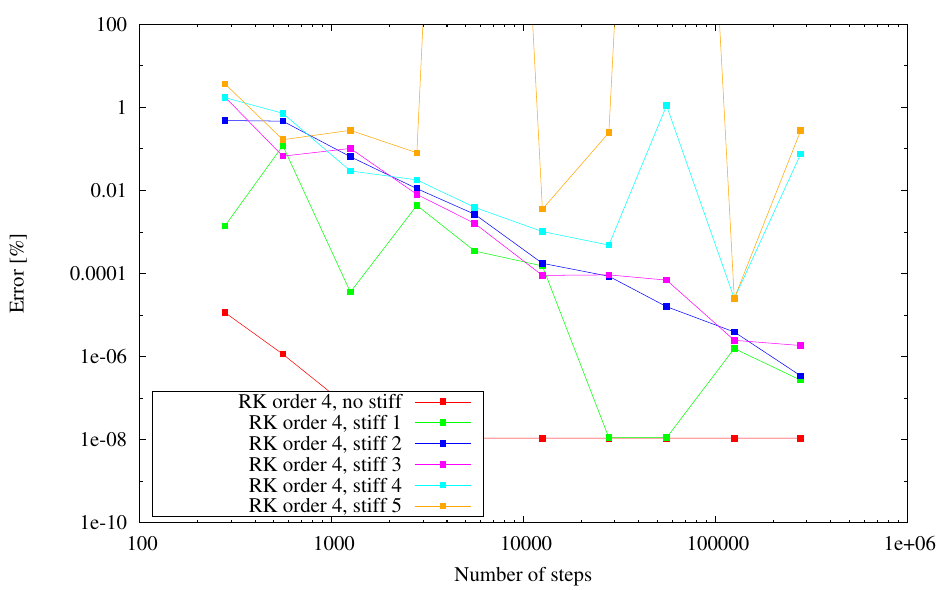} }}%
    \caption{Relative error for various stiffness levels and integration step sizes of the simulated T-BMT equation.}%
    \label{fig:stiffness}%
\end{figure*}

When hard edge models are encountered, stiffness becomes the bottleneck of precision tracking. When solving the equation of motion and/or the T-BMT equation, some singularities may arise at the entrance/exit of the hard edge elements due to discontinuities in the fields calculated or their derivatives. In order to assess its impact on the convergence of the Runge Kutta tracker, one simulated several cases in Mathematica where one started from a smooth model with continuous functions and approximate it by stepwise functions for which the stiffness can be controlled. For instance, increasing the stiffness can be achieved by reducing the width of the stepwise function, therefore increasing the slope at the entrance/exit of the element (the hard edge model corresponds to an infinite slope). This is better illustrated in fig \ref{stiff}.
The idea is to investigate the convergence of the Runge Kutta tracker by slowly increasing the stiffness of the system. One first starts from smooth functions describing the spin precession components as sinusoidal oscillations. Then, using the above model, one can vary the stiffness of the system: from one case to the next, one reduces by half the rising time of the spin precession components. The T-BMT equation is then solved for each case using a specific order and a fixed step size of the explicit Runge Kutta tracker in Mathematica. The relative error after 10 revolution periods is computed for each case corresponding to a fixed integration step size, a specific order of the Runge Kutta method and a given spin precession function. In one case the assumed reference solution corresponds to the $9th$ order Runge Kutta method with the smallest step size. In the second case, the assumed reference solution corresponds to the second order analytical solution. The results are displayed in fig \ref{fig:stiffness} where one can observe some oscillatory behavior of the relative error due to instability of the Runge Kutta method. However, in general, the higher the stiffness, the larger the error becomes for a given step size. Nevertheless both the analytical and the Runge Kutta solutions converge towards each other. Using implicit Runge Kutta tracker does reduce the error. This will be discussed in later publication.
 
\end{document}